\documentclass{optica-article}

\journal{opticajournal} % for journals or Optica Open

\articletype{Research Article}
\usepackage{xspace}
\usepackage{stfloats}
\usepackage{graphicx}
\usepackage{float}
\usepackage{subfigure}
\usepackage{multirow}
\begin{document}

\title{Monte-Carlo based non-line-of-sight underwater wireless optical communication channel modeling and system performance analysis under turbulence}

\author{Peng Yue,\authormark{1} XiangRu Wang,\authormark{1,*} Shan Xu,\authormark{1} and YunLong Li\authormark{1}}

\address{\authormark{1}State Key Laboratory of Integrated Service Networks, Xidian University, Xian 710071, China }

\email{\authormark{*}22011211032@stu.xidian.edu.cn} %% email address is required; see note below about the corresponding author designation

% use {asbstract*} to suppress the copyright line. Copyright information will be added in production

\begin{abstract*} 
Compared with line-of-sight (LOS) communication, nonline-of-sight (NLOS) underwater wireless optical communication (UWOC) systems have garnered extensive attention because of their heightened suitability for the intricate and dynamic underwater environment. In the NLOS channel, photons can reach the receiver by sea surface reflection or particle scattering. However, research lacks comprehensive channel models that incorporate sea surface reflection and particle scattering. Moreover, the presence of ocean turbulence introduces random fluctuations in the received optical signal based on the average light intensity. Consequently, this paper adopts the Monte Carlo simulation method (MCS) to solve the fading-free impulse response of the joint reflection-scattering channel. Furthermore, a weighted double gamma function (WDGF) is proposed to characterize the channel impulse response (CIR). Based on the closed CIR model, the average bit error rate and the performance of the interruption probability of the UWOC system under turbulence are analyzed. The conclusions obtained are intended to assist in the design and performance evaluation of NLOS UWOC systems.

\end{abstract*}

%%%%%%%%%%%%%%%%%%%%%%%%%%  body  %%%%%%%%%%%%%%%%%%%%%%%%%%
\section{Introduction}
In recent years, the demand for marine surveys, climate monitoring, ecological protection, and other activities has caused the development of underwater information transmission technology to be widely concerned by researchers in various countries. Underwater wireless optical communication (UWOC) has become a very potential and attractive communication method due to its advantages of high communication rate and low latency\cite{1,2,3,new1,new2,new3}. However, a complex underwater environment poses significant challenges to high-quality communication, such as signal interruptions due to absorption and scattering caused by inorganic particles, zooplankton, and organic debris. Therefore, overcoming the problem of underwater fading and establishing a reasonable channel model are key issues that need to be urgently addressed for the further development of underwater communication.

In view of the fact that light propagates linearly in a homogeneous medium, previous studies of UWOC channels and systems have focused on line-of-sight (LOS) link configurations. In this case, the photons reach the receiver directly through the line-of-sight link and are suitable for point-to-point communication. When the transceiver cannot be strictly aligned due to water fluctuations, the probability of system outage will increase significantly. Compared with the LOS configuration, the Non-Line-of-Sight (NLOS) configuration relaxes the requirement of strict transceiver alignment, can effectively deal with the transceiver straight-line distance obstacle, and has excellent system stability, which is very suitable for the complex and changing underwater environment.

The photons propagating in the NLOS channel mainly rely on particle scattering to reach the receiver, which is called the scattering channel. The scattering order is related to seawater conditions and transceiver spacing. Multiple scattering will occur when the particle density is large and the propagation distance is wide. To investigate the propagation characteristics of light for UWOC links, researchers have proposed and evaluated various channel models analytically and numerically. In theoretical analysis of the pulse response, vector radiative transfer (VRT) is typically used to analyze the attenuation and multiple scattering effects. Refs. \cite{4,5} presented the modeling and performance evaluation of a UWOC channel by accurately solving the VRT equation. However, these studies have several limitations. For instance, the receiving aperture was not taken into consideration \cite{4}, and the analytical results are obtained after introducing a series of hypotheses and approximations to simplify the VRT equation \cite{5}, which degrades the accuracy of the model. To address these issues, a Monte Carlo approach was proposed in Ref. \cite{6} to investigate the temporal impulse response of UWOC systems. The Monte Carlo method is based on a series of photon-particle collisions to simulate ocean radiation transport and quantifies the impulse response by statistical analysis of the received photons, which has been widely applied to radiative transfer in scattering media. Refs. \cite{6,7,8} investigated the channel impulse response (CIR) of multiple channels in dispersion under LOS communication links using Monte Carlo simulation (MCS), and the results showed that the channel time dispersion is negligible for moderate propagation distance (< 100 m) except for highly turbid seawater. However, all of these reported works failed to provide an accurate mathematical model of the CIR. To remedy this limitation, Tang et al. proposed a double Gamma function to fit the CIR results obtained from MCS \cite{9,10}, and obtained a better fit under turbid seawater. Subsequently, a combination of exponential and arbitrary power functions was proposed to fit the CIR, along with a more reasonable explanation of the channel absorption and scattering behavior in Ref. \cite{11}.To explain the percentage contribution of LOS light and scattered light to the received power, see Ref. \cite{12} proposed a CIR model based on the superposition of one impulsive component and one dispersive component.

Unfortunately, the above channel model is only applicable when the transceivers are strictly aligned and the transmitter's center axis is perpendicular to the receiver's aperture. In real seawater environments, obstacles and turbulent interference can render the UWOC system ineffective. Given these limitations, reflective non-line-of-sight (NLOS) UWOC systems have been proposed \cite{13}. In reflective NLOS implementations, the transceiver utilizes the reflection of the sea surface to overcome link obstacles, thus avoiding the blockage of the propagation path. Consequently, reflective NLOS channel characteristics are closely connected with the sea surface, and related research has been conducted in the literature. Refs. \cite{13,14,15} assumed a stable sea surface. A mathematical model of the NLOS channel was built considering link attenuation, sea-air interface reflections, and receiver field of view (FOV) \cite{13,14}. MCS method was adopted in \cite{15} to track the propagation trajectory of photons, and the impacts of different water types on the NLOS channel were studied. Compared with the LOS link, the channel bandwidth of the reflective link is much smaller than that of the LOS link under the same seawater conditions. Ref. \cite{16} derived the path loss characteristics theoretically under a single specular reflection channel in non-turbid seawater, obtained the highest channel gain of 3 dB, and carried out water tank experiments to verify the results. However, the ideal assumption of a stable sea surface and negligible scattering may differ from the realistic channel. Refs. \cite{17,18} modeled the sea surface slope as a Gaussian distributed random variable while taking into account the effect of particle scattering, and analyzed the channel path loss for the horizontal and vertical links, respectively. The results indicate that the received signal is severely degraded under the reflective channel, but the scattering light may dimmish the negative effects. Therefore, the construction of a combined surface reflection and scattering channel is positive for the construction of a reliable UWOC system. However, there are no closed-form functions to model the CIR for the joint channel with reflection and scattering effects in the current research work.

In addition to absorption scattering, ocean turbulence is one of the factors affecting the quality of underwater communications. Turbulence is caused by water temperature salinity ups and downs, which can cause light intensity to flicker at the receiving end. Normalized light intensity models in the presence of turbulence have been well studied. Channel characterization is ultimately studied to build the communication system and analyze the system's performance. Literature has been developed to carry out theoretical studies on the performance of UWOC systems. Ref. \cite{19} analyzes the average bit error rate (BER) performance in the presence of turbulence by considering the absorption and scattering effects of seawater. Ref. \cite{20} proposes a multiple-input multiple-output (MIMO) system based on scattering channels and analyzes the average BER of the system in the presence of turbulence. However, the channels considered in the above communication systems are absorbing scattering channels and the effect of sea surface reflection on the system is ignored.

To make up for the deficiencies of the existing research, in this paper, we first propose a joint channel of rough sea surface reflection and particle scattering, design a Monte Carlo simulation algorithm, and propose a simple expression for the weighted dual Gamma function to fit the CIR obtained from the simulation. Based on the results, the average BER and outage probability of the UWOC system in the presence of turbulence are analyzed. To the best of our knowledge, this is the first time that a mathematical expression for the channel CIR that integrates reflection and scattering effects is presented, which helps to characterize the joint channel properties of particle scattering and rough surface reflection.

The main contributions of this paper are as follows:

1. From the scattering and reflection mechanism, a Monte Carlo simulation algorithm for the joint reflection and scattering channel is provided to obtain the CIR of the channel.

2. A weighted double Gamma function is proposed for fitting the CIR simulation results and obtaining a better fit under coastal and turbid water conditions.

3. The average BER and outage probability performance of UWOC systems under joint channels is analyzed based on the closed CIR model and the lognormal turbulence model.

This paper is organized as follows: Section 2 describes the absorption and scattering characteristics of seawater, the stochastic sea surface model, and the turbulence model, while Section 3 outlines the details of MCS and our proposed closed-form CIR model, and Section 4 provides The BER and outage probability analysis methods. Numerical results and discussion will be presented in Section 5, followed by conclusions in Section 6.

\section{Channel and system model}

\subsection{Optical characterization of seawater}

Absorption and scattering are the main reasons for the loss of light energy by seawater. The former is an irreversible process in which photons lose heat through interaction with water molecules and soluble organic matter. Scattering is the attenuation of energy in the original direction of motion of photons due to the deviation of the beam from its original path caused by changes in the density of suspended particles or seawater. The strength of the scattering effect is determined by the various particles present underwater, so that the effect on photons is more pronounced in coastal areas than in the open sea. The energy loss caused by the above two effects can be evaluated by absorption coefficient $a\left( \lambda  \right)$ and scattering coefficient $b\left( \lambda  \right)$, respectively. The attenuation coefficient  $c\left( \lambda  \right)=a\left( \lambda  \right)+b\left( \lambda  \right)$ describes the overall energy loss in the underwater channel. The typical value of $a\left( \lambda  \right)$, $b\left( \lambda  \right)$ and $c\left( \lambda  \right)$ in turbid sea water are shown in Table \ref{tab:1}.

\begin{table}[htbp]
\centering
\caption{\bf Extinction Coefficients for Different Types of Water}
\begin{tabular}{cccc}
\hline
Water Type & $a(\lambda )[{{m}^{-1}}]$ & $b(\lambda )[{{m}^{-1}}]$ & $c(\lambda )[{{m}^{-1}}]$ \\
\hline
Costal ocean & $0.178$ & $0.220$ & $0.398$ \\
Turbid Ocean & $0.295$ & $1.875$ & $2.17$ \\
\hline
\end{tabular}
  \label{tab:1}
\end{table}

Unlike atmospheric optical links, underwater scattering effects are complicated by the presence of inorganic matter, dissolved salts, and organic particles. Photon scattering effects are more pronounced and of higher scattering order in turbid waters. To analyze the scattering properties of seawater for photons, the scattering phase function (SPF) $\beta \left( \theta ,\lambda  \right)$ is introduced to describe the energy distribution at the scattering angle $\theta $ with the expression is introduced to calculate the energy distribution of the scattering angle with the expression as:
\begin{equation}
\label{eq1}
    2\pi \int_{0}^{\pi }{\beta \left( \theta ,\lambda  \right)}\sin \theta d\theta =1
\end{equation}

The original value of SPF was measured by Petzold in the optical band centered at 514 nm \cite{21}. Based on the experimental measurements, scholars have proposed several typical closed-form models to characterize the scattering properties of water bodies. Among them, the Henyey-Greenstein phase function (HGPF) is widely used because of the simple form with:
\begin{equation}
\label{eq2}
    {{\beta }_{HG}}(\theta )=\frac{1-{{g}^{2}}}{4\pi {{\left( 1+{{g}^{2}}-2g\cos (\theta ) \right)}^{3/2}}}
\end{equation}
where g is the average cosine of $\theta $. It is convenient for solution of $\theta $, but the difference between HGPF and Petzold’s measurements is not negligible in small forward angles $\left( \theta <20{}^\circ  \right)$ and backward angles $\left( \theta <130{}^\circ  \right)$. Therefore, two-term HG phase function (TTHGPF) which is the linear combinations of HG can upgrade the fitting performance of large and small angles. 

\begin{figure}[htbp]
\centering\includegraphics[width=8cm]{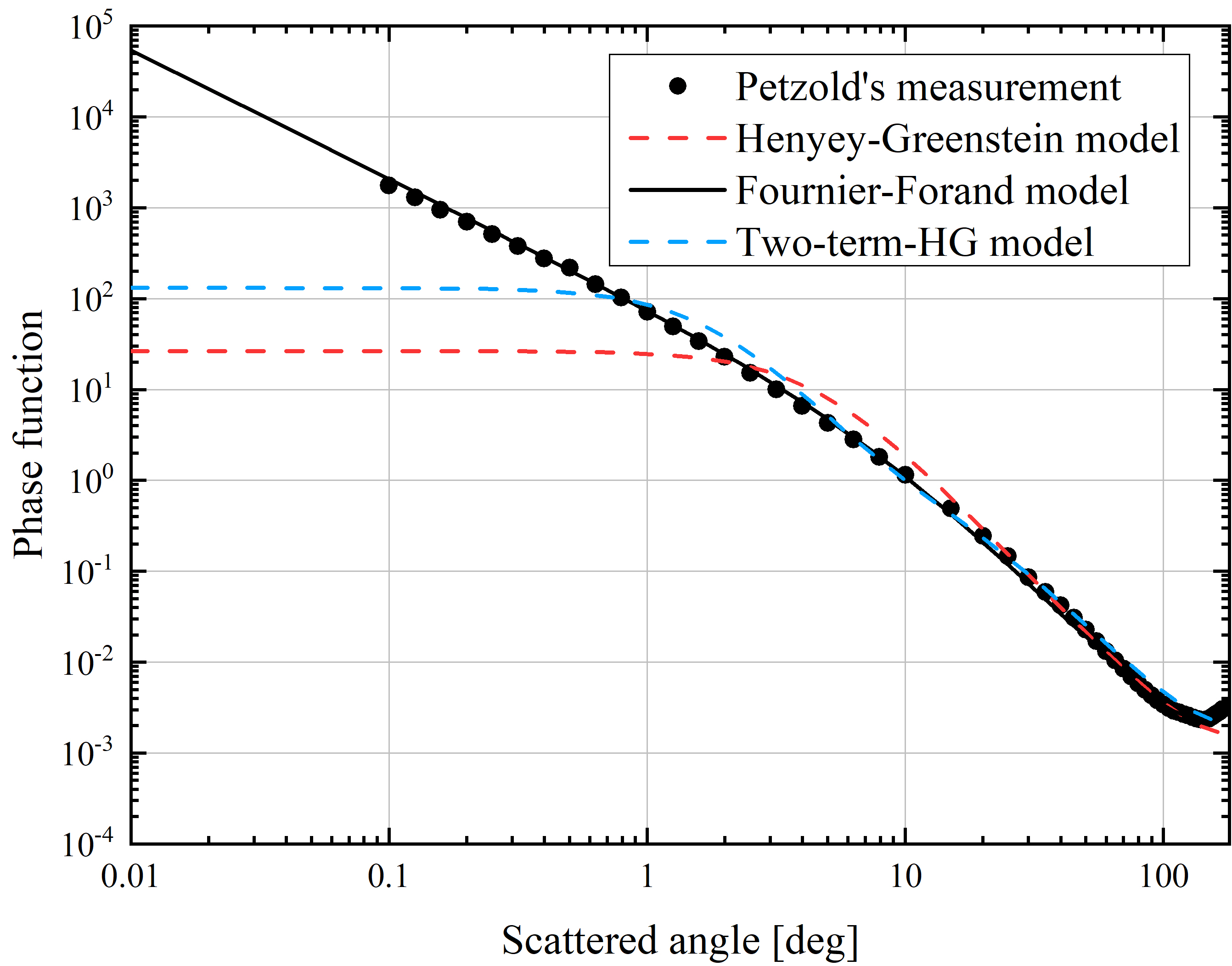}
\caption{Comparison of different phase functions.}
\label{fig:1}
\end{figure}

However, the value of such a model is still smaller than Petzold’s measurement for  $\theta <1{}^\circ $.
A more accurate model Fournier-Forand phase function (FFPF) is proposed as an alternative to HGPF. FFPF is obtained by applying the anomalous diffraction and Mie scattering approximations. The expression is given by \cite{21}:
\begin{equation}
\label{eq3}
    \begin{aligned}
         & {{\beta }_{FF}}(\theta )=\frac{1}{4\pi {{\left( 1-\delta  \right)}^{2}}{{\delta }^{\upsilon }}}\left\{ \upsilon \left( 1-\delta  \right)-\left( 1-{{\delta }^{\upsilon }} \right) \right. \\ 
 & \quad \quad \quad \quad +\left. \left[ \delta \left( 1-{{\delta }^{\upsilon }} \right)-\upsilon \left( 1-\delta  \right) \right]{{\sin }^{-2}}\left( \theta /2 \right) \right\} \\ 
 & \quad \quad \quad \quad +\frac{1-\delta _{180}^{\upsilon }}{16\pi \left( {{\delta }_{180}}-1 \right)\delta _{180}^{\upsilon }}\left( 3{{\cos }^{2}}\left( \theta  \right)-1 \right) \\ 
    \end{aligned}
\end{equation}
where $\upsilon =\left( 3-\mu  \right)/2$ and $\delta =4/3{{\left( n-1 \right)}^{2}}{{\sin }^{2}}\left( \theta /2 \right)$, $n$ is the refractive index of the water, and ${{\delta }_{180}}$ is $\delta$ evaluated at $\theta =180{}^\circ$. Fig. \ref{fig:1} shows the HGPF, TTHGPF, the FFPF and Petzold’s measured data for comparison. We can see that FFPF fits better to the measurements for arbitrary scattering angles. Therefore, we adopt the FFPF as the scattering phase function.

As shown in Eq.(\ref{eq3}), the complex expression makes it challenging to solve for $\theta $ directly. We develop a numerical solution for scattering angle by establishing a correspondence between $\theta $ and the random number \cite{22}. The steps consist of the following three parts:
(1). We divide the scattering angle in the interval $\left[ 0,\pi  \right]$ into $N$ discrete uniform points. It is effortless to calculate the corresponding SPF values from Eq.(\ref{eq3}). According to Eq.(\ref{eq1}), the phase function normalization formula can be rewritten as:
\begin{equation}
\label{eq4}
    2\pi \sum\limits_{m=1}^{N}{{{\beta }_{FF}}\left( {{\theta }_{m}} \right)}\sin {{\theta }_{m}}\Delta {{\theta }_{m}}=1
\end{equation}
(2). For each scattering angle ${{\theta }_{i}},i=1,2,\ldots ,N$, its relation to the random number is built by:
\begin{equation}
\label{eq5}
    {{s}_{i}}=2\pi \sum\limits_{m=1}^{i}{{{\beta }_{FF}}\left( {{\theta }_{m}} \right)}\Delta {{\theta }_{m}}.
\end{equation}
(3). Finally, choosing a uniformly distributed random variable ${{\xi }_{i}}$ in $\left[ 0,1 \right]$, if its value satisfies the condition ${{s}_{i-1}}<{{\xi }_{i}}\le {{s}_{i}}$, the scattering angle is determined as ${{\theta }_{i}}$.

\subsection{Random sea surface model}

\begin{figure}[htbp]
\centering\includegraphics[width=8cm]{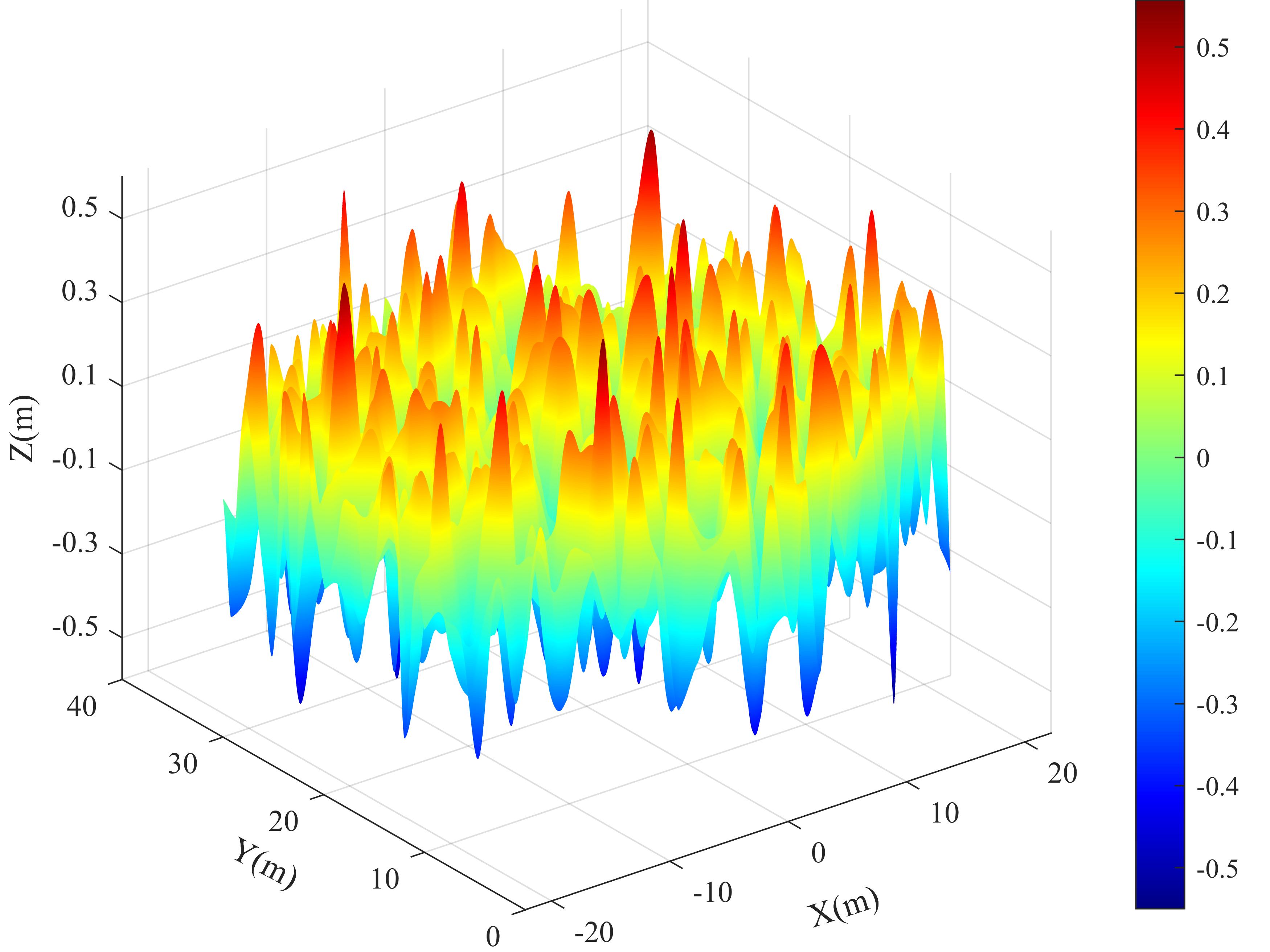}
\caption{Model of the sea surface as a hexagonal grid of triangular wave facets.}
\label{fig:2}
\end{figure}

Assuming the wind blows over the water, at a fixed horizontal position, the height of the surface varies randomly due to the passage of waves. To quantify a wind-driven sea surface, we used the capillary waves model proposed by Preisendorfer and Mobley \cite{23}. The core concept of the model is dividing the undulating sea surface into a series of similar triangular facets, called “triads”. Each facet can be treated as a local plane. The reflection and transmission of light on different facets obey the laws of geometric optics. Fig. \ref{fig:2} shows the capillary waves model.

The creation of a random sea surface starts with a flat hexagonal grid consisting of several similar triads. The elevation of each point follows a Gaussian distribution $N\left( 0,\sigma _{\eta }^{2} \right)$. The connection of the three points with random elevations in a triad defines one capillary wave facet. The realization of random surfaces is finished by constructing all facets above or below the hexagonal grid. The upwind and crosswind dimensions of a triad $\left( \delta ,\varepsilon  \right)$, and the variance $\sigma _{\eta }^{2}$ are given by \cite{24}:
\begin{equation}
\label{eq6}
    \begin{aligned}
  & \delta =1 \\ 
 & \frac{{{\varepsilon }^{2}}}{{{\delta }^{2}}}=\frac{3{{a}_{u}}}{4{{a}_{c}}} \\ 
 & \sigma _{\eta }^{2}=\frac{1}{2}{{a}_{u}}{{\delta }^{2}}U, \\ 
\end{aligned}
\end{equation}
where ${{a}_{u}}$ and ${{a}_{c}}$ are constants found experimentally by Cox and Munk \cite{25} with the expression as:
\begin{equation}
\label{eq7}
    \begin{aligned}
  & {{a}_{u}}=3.16\times {{10}^{-3}}s/m \\ 
 & {{a}_{c}}=1.92\times {{10}^{-3}}s/m \\ 
\end{aligned}
\end{equation}
and $U$ is the wind speed in meters per second which is measured at an anemometer height 12.5m above the mean sea level. Based on the above definitions, the slopes of any facet follows Gaussian distribution:
\begin{equation}
\label{eq8}
    p\left( {{z}_{x}},{{z}_{y}} \right)=\frac{1}{2\pi {{\sigma }_{u}}{{\sigma }_{c}}}\exp \left[ -\frac{1}{2}\left( \frac{z_{x}^{2}}{\sigma _{u}^{2}}+\frac{z_{y}^{2}}{\sigma _{c}^{2}} \right) \right]
\end{equation}
where ${{z}_{x}}$ and ${{z}_{y}}$ represents the slopes of a facet with ${{z}_{x}}={\partial z}/{\partial x}\;$ and ${{z}_{y}}={\partial z}/{\partial y}\;$. $\sigma _{u}^{2}$ and $\sigma _{c}^{2}$ are the isotropic variance of slopes given by:
\begin{equation}
\label{eq9}
    \begin{aligned}
  & \sigma _{u}^{2}={{a}_{u}}U \\ 
 & \sigma _{c}^{2}={{a}_{c}}U. \\ 
\end{aligned}
\end{equation}
Note that we ignore the upwind and crosswind skewness in this paper for simplicity. Then the probability distribution of the directions of the normal of the facet can be expressed as follows:
\begin{equation}
\label{eq10}
    p\left( {{\theta }_{n}},{{\phi }_{n}} \right)=\frac{1}{2\pi {{\sigma }_{u}}{{\sigma }_{c}}}\exp \left[ -\frac{{{\tan }^{2}}{{\theta }_{n}}}{2{{\sigma }_{u}}{{\sigma }_{c}}} \right]\tan {{\theta }_{n}}{{\sec }^{2}}{{\theta }_{n}}
\end{equation}
where ${{\theta }_{n}}$ and ${{\phi }_{n}}$ are the zenith and azimuth angle of the normal of the facet, respectively. 	

\subsection{Fading statistic of turbulence}

Optical turbulence arises from fluctuations in water temperature and salinity, leading to random changes in refractive index \cite{26}. In the previous section, we introduced the seawater absorption and scattering properties, along with the rough sea surface model. These components can be represented as the fading-free impulse response (FFIR) ${{h}_{0}}(t)$ by Monte Carlo simulation methods. To incorporate turbulence effects, we multiply ${{h}_{0}}(t)$ by a multiplicative fading coefficient $\tilde{h}$, which accounts for the stochastic fading of the optical signal caused by turbulence, following a lognormal distribution. Lognormal distribution turbulence statistical models are widely used to characterize weak turbulence with scintillation coefficients $\sigma _{I}^{2}<1$, due to their simplicity of form. Based on this, the PDF of the channel fading coefficient $\tilde{h}=\exp \left( 2X \right)$ is expressed as:
\begin{equation}
\label{eq11}
    {{f}_{{\tilde{h}}}}\left( {\tilde{h}} \right)=\frac{1}{2\tilde{h}\sqrt{2\pi \sigma _{X}^{2}}}\exp \left( -\frac{{{\left( \ln \left( {\tilde{h}} \right)-2{{\mu }_{X}} \right)}^{2}}}{8\sigma _{X}^{2}} \right)
\end{equation}
where $X=1/2\ln \left( {\tilde{h}} \right)$ indicates the fading log-amplitude obeying a Gaussian distribution with mean ${{\mu }_{X}}$ and variance $\sigma _{X}^{2}$. Ref.\cite{27} verifies the fact that there exists such an equational relationship ${{\mu }_{X}}=-\sigma _{X}^{2}$ between the mean and variance. Thus, the log-normal distribution is a function with a single parameter. To further calculate the mean value of the fading factor distribution, the relationship between the scintillation index and the fading factor needs to be clarified. Under weak turbulence conditions, the scintillation index of plane and spherical waves can be calculated as \cite{28}:
\begin{equation}
\label{eq12}
    \begin{aligned}
  & \sigma _{I}^{2}=8{{\pi }^{2}}k_{0}^{2}{{d}_{0}}\int_{0}^{1}{\int_{0}^{\infty }{\kappa {{\Phi }_{n}}}}\left( \kappa  \right) \\ 
 & \ \quad \ \ \ \ \times \left\{ 1-\cos \left[ \frac{{{d}_{0}}{{\kappa }^{2}}}{{{k}_{0}}}\xi \left( 1-(1-\Theta )\xi  \right) \right] \right\}d\kappa d\xi  \\ 
\end{aligned}
\end{equation}
where plane and spherical waves correspond to $\Theta =1$ and $0$, respectively. ${{k}_{0}}=2\pi /\lambda $ and $\kappa $ is wave number and scalar spatial frequency, respectively. ${{\Phi }_{n}}\left( \kappa  \right)$ is the power spectrum of turbulent fluctuations given by \cite{29}. In the weakly turbulent case considered in this paper, the relationship between the scintillation coefficient and the lognormal variance is as follows: $\sigma _{X}^{2}=0.25\ln \left( 1+\sigma _{I}^{2} \right)$ \cite{20}.

\section{Fading-Free Impulse response modeling}

\subsection{Monte-Carlo Simulation}

As a typical model for exposing the propagation laws of light underwater, the radiative transfer equation (RTE) can be solved analytically and numerically. One of the most commonly used numerical methods is Monte Carlo simulation. By generating a large number of photons and then simulating the interactions of each photon with the medium, the channel characteristics are ultimately evaluated. Compared with the analytical model Monte-Carlo method is flexible for various system geometries and does not impose any restrictions on the propagation of photons. Therefore, MCS is applied here to evaluate the CIR.
\begin{figure}[htbp]
\centering\includegraphics[width=8cm]{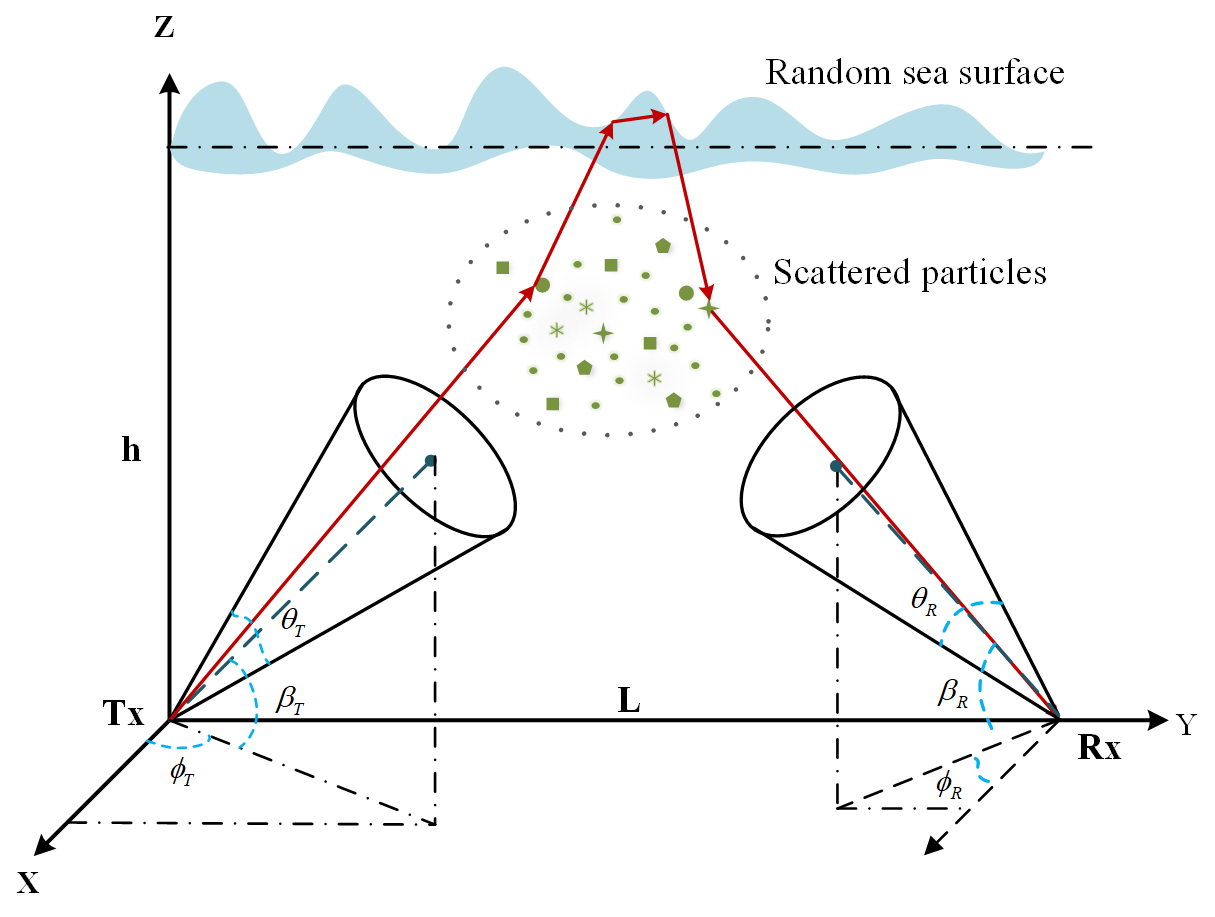}
\caption{Geometric setup of the system.}
\label{fig:3}
\end{figure}

Fig.\ref{fig:3} shows the geometry of the NLOS link considering both reflection and scattering caused by random sea surface and seawater, respectively. The transmitter Tx and receiver Rx of aperture area ${{A}_{r}}$ are placed on the seabed below the sea level with depth $h$. Tx and Rx are located at $\left( 0,0,0 \right)$ and $\left( 0,L,0 \right)$ respectively. Denote the Tx beam fullwidth divergence by ${{\theta }_{T}}$ and Rx FOV by ${{\theta }_{R}}$. We term ${{\phi }_{T}}$ and ${{\phi }_{R}}$ as the azimuth angle at transmitter and receiver respectively. To locate beam initial elevation angles are considered as ${{\beta }_{T}}$ and ${{\beta }_{R}}$.

The MCS starts with generating numerical photons. The basic properties of each photon are uniquely determined by spatial position $\left( {{x}_{i}},{{y}_{i}},{{z}_{i}} \right)$, direction vector $\vec{p}$ defined by elevation and azimuth angles and weight ${{W}_{i}}$. The generated photon may be reflected by the sea surface or scattered by the water more than once, after every interaction with a particle or surface, the photon updates direction and weight, till it is received or lost. To facilitate the estimation of the channel impact response, we refer to the particle reflection and rough surface scattering events collectively as directional deflection events. Assume that a photon may undergo M-order directional deflection events from the time it is emitted to the time it is captured by the receiver. By simulation the propagation of all photons, we obtain the CIR.

Initially, each photon with unity weight is emitted from the origin of the coordinate system. The initial emitting process is regarded as the zero-order direction deflection. Each photon’s initial direction must be confined within the full beam angle ${{\theta }_{T}}$ with solid angle $2\pi \left( 1-\cos {{\theta }_{T}}/2 \right)$. Denote the initial direction of an individual photon concerning the elevation angle as ${{\theta }_{0}}$, and concerning the azimuth angle as ${{\varphi }_{0}}$. For a uniformly distributed light source, the emission angles are set as: 
\begin{equation}
\label{eq13}
    \begin{aligned}
  & {{\theta }_{0}}={{\beta }_{T}}-{{\theta }_{T}}/2+{{\theta }_{T}}{{\xi }^{\left( \theta  \right)}} \\ 
 & {{\varphi }_{0}}=2\pi {{\xi }^{\left( \varphi  \right)}} \\ 
\end{aligned}
\end{equation}
where ${{\xi }^{\left( \theta  \right)}}$ and  ${{\xi }^{\left( \varphi  \right)}}$ are two independent random variables uniformly distributed between 0 and 1. For a non-uniform source, the value of ${{\xi }^{\left( \theta  \right)}}$ is set according to the angular intensity distribution of the light source. Before interaction with a particle or sea surface photon will travel a random distance $\Delta s$ called step size, which can be expressed as\cite{21}:
\begin{equation}
\label{eq14}
    \Delta s=-\frac{\ln \left( {{\xi }^{\left( s \right)}} \right)}{b\left( \lambda  \right)}
\end{equation}
where ${{\xi }^{\left( s \right)}}$ is a uniformly distributed random variable in the interval of [0,1]. Then the position of a photon can be determined according to the direction of photon emission and the random step $\Delta s$. We describe the position’s spatial position with three Cartesian coordinates and the migration direction with three direction cosines from two direction angles $\theta$ and $\varphi$. The initial direction cosines are specified by $\mu _{x}^{0}=\cos {{\theta }_{0}}\cos {{\varphi }_{0}}$, $\mu _{y}^{0}=\cos {{\theta }_{0}}\sin {{\varphi }_{0}}$, $\mu _{z}^{0}=\sin {{\theta }_{0}}$. For a photon located at $\left( {{x}_{i}},{{y}_{i}},{{z}_{i}} \right)$ traveling a distance $\Delta s$ in the direction $\left( \mu _{x}^{i},\mu _{y}^{i},\mu _{z}^{i} \right)$, its coordinates are updated according to the three-dimensional geometry by:
\begin{equation}
\label{eq15}
    \begin{aligned}
  & {{x}_{i+1}}={{x}_{i}}+\mu _{x}^{i}\Delta s, \\ 
 & {{y}_{i+1}}={{y}_{i}}+\mu _{y}^{i}\Delta s, \\ 
 & {{z}_{i+1}}={{z}_{i}}+\mu _{z}^{i}\Delta s. \\ 
\end{aligned}
\end{equation}

Both surface reflection and particle scattering can lead to the deflection of the photon direction. We distinguish between these two cases according to the relative spatial position of the photon to the sea level. In simple terms, if the current position of the photon is below sea level, the photon undergoes scattering based on the particle, otherwise reflection based on rough sea surface occurs.

When interacting with the particles, the photon’s migration direction can be characterized by FFPF, i.e., Eq.(\ref{eq3}). Then the deviation from the current direction is determined by a pair of angles $\left( \theta ,\varphi  \right)$, here $\theta$ is the zenith angle calculated by Eq.(\ref{eq5}) and $\varphi =2\pi \xi$ is azimuth angle with uniform distribution in the interval of $\left[ 0,1 \right]$. After scattering from the direction $\left( \mu _{x}^{i},\mu _{y}^{i},\mu _{z}^{i} \right)$, the new direction cosines are calculated by $\left( {{u}_{x}},{{u}_{y}},{{u}_{z}} \right)$:
\begin{equation}
\label{eq16}
    \begin{aligned}
  & \mu _{x}^{i+1}=\frac{\sin \theta }{\sqrt{1-{{\left( \mu _{x}^{i} \right)}^{2}}}}\left( \mu _{x}^{i}\mu _{z}^{i}\cos \varphi -\mu _{y}^{i}\sin \varphi  \right)+\mu _{x}^{i}\cos \theta , \\ 
 & \mu _{y}^{i+1}=\frac{\sin \theta }{\sqrt{1-{{\left( \mu _{x}^{i} \right)}^{2}}}}\left( \mu _{y}^{i}\mu _{z}^{i}\cos \varphi -\mu _{z}^{i}\sin \varphi  \right)+\mu _{y}^{i}\cos \theta , \\ 
 & \mu _{z}^{i+1}=-\sin \theta \cos \varphi \sqrt{1-{{\left( \mu _{z}^{i} \right)}^{2}}}+\mu _{z}^{i}\cos \theta . \\ 
\end{aligned}
\end{equation}

Another key event is the interaction between photon and random sea surface. When the photon strike the sea surface modeled by a random facet, the direction of a facet normal can be obtained by:
\begin{equation}
\label{eq17}
    \begin{aligned}
  & {{\phi }_{n}}=2\pi {{\xi }^{\left( {{\phi }_{n}} \right)}}, \\ 
 & {{\theta }_{n}}=\arctan \sqrt{-2{{\sigma }_{u}}{{\sigma }_{c}}\ln \left( \frac{2-{{\xi }^{\left( {{\theta }_{n}} \right)}}}{2} \right)}. \\ 
\end{aligned}
\end{equation}
here, ${{\xi }^{\left( {{\phi }_{n}} \right)}}$ and ${{\xi }^{\left( {{\theta }_{n}} \right)}}$ are two independent random variables uniformly distributed between zero and one. After determining the direction of normal, the direction of a reflected photon can be calculated from the incident direction $\left( {{\mu }_{x}},{{\mu }_{y}},{{\mu }_{z}} \right)$ and facet normal based on Snell’s Law:
\begin{equation}
\label{eq18}
    \begin{aligned}
  & {{\mu }_{x,r}}={{u}_{x}}-2\left( {{\mu }_{x}}\cos {{\theta }_{n}}\cos {{\phi }_{n}}+{{\mu }_{y}}\cos {{\theta }_{n}}\sin {{\phi }_{n}}+{{\mu }_{z}}\sin {{\theta }_{n}} \right)\left( \cos {{\theta }_{n}}\cos {{\phi }_{n}} \right), \\ 
 & {{\mu }_{y,r}}={{u}_{y}}-2\left( {{\mu }_{x}}\cos {{\theta }_{n}}\cos {{\phi }_{n}}+{{\mu }_{y}}\cos {{\theta }_{n}}\sin {{\phi }_{n}}+{{\mu }_{z}}\sin {{\theta }_{n}} \right)\left( \cos {{\theta }_{n}}\sin {{\phi }_{n}} \right), \\ 
 & {{\mu }_{z,r}}={{u}_{z}}-2\left( {{\mu }_{x}}\cos {{\theta }_{n}}\cos {{\phi }_{n}}+{{\mu }_{y}}\cos {{\theta }_{n}}\sin {{\phi }_{n}}+{{\mu }_{z}}\sin {{\theta }_{n}} \right)\sin {{\theta }_{n}}. \\ 
\end{aligned}
\end{equation}

Each reflected photon should be weighted by the reflection coefficient that is given as \cite{17}:
\begin{equation}
\label{eq19}
    R=\left\{ \begin{array}{*{35}{r}}
   \frac{1}{2}{{\left[ \frac{\tan ({{\theta }_{t}}-{{\theta }_{in}})}{\tan ({{\theta }_{t}}+{{\theta }_{in}})} \right]}^{2}}+\frac{1}{2}{{\left[ \frac{\sin ({{\theta }_{in}}-{{\theta }_{t}})}{\sin ({{\theta }_{in}}+{{\theta }_{t}})} \right]}^{2}},{{\theta }_{in}}<{{\theta }_{c}}  \\
   1\ \ \quad \quad \quad \quad \quad \quad \quad ,{{\theta }_{in}}\ge {{\theta }_{c}}  \\
\end{array} \right.
\end{equation}
where ${{\theta }_{in}}$ is the angle between of incident photon and surface normal, and ${{\theta }_{t}}$ is the angle between the reflected photon and surface normal. ${{\theta }_{c}}=\arctan \left( {{n}_{a}}/{{n}_{w}} \right)$ is the critical angle of the interface with ${{n}_{a}}$ and ${{n}_{w}}$ as the refractive index of air and seawater, respectively. The weight of the reflected photon is updated as:
\begin{equation}
\label{eq20}
    {{W}_{r}}=R{{W}_{before}}
\end{equation}
where ${{W}_{before}}$ is the weight of the photon before reflection.

It is worth noting that the rough sea surface may cause the photon to hit the sea surface again, which is referred to as the multiple reflection process in Ref.\cite{17}. When the photon still has an upward reflection direction after hitting the sea surface, we consider that such photons will be reflected again with the sea surface. In Section 2, we create rough sea surfaces by dividing the sea surface into multiple capillary wave facets with different slopes to study the multiple reflection behavior of photons. The spatial resolution of these wavefronts is at the centimeter level, which is smaller enough than the link range, so we neglect the trajectories among the multiple reflection events. Therefore, we assume that a photon can hit the sea surface again immediately during the multiple reflection process without internal trajectories and only reconsider its directions. If the direction of the photon is downward after experiencing multiple reflections, we assume that the photon is reflected into the water and is scattered by the particle after moving a random step in the current direction.

 For a photon that falls within the receiver FOV, we consider it to be captured by the receiver. The probability of a photon reaching the receiver after m scattering events can be calculated by integrating the scattering phase function ${{\beta }_{FF}}(\theta )$ over an appropriate set of angles, written as:
 \begin{equation}
 \label{eq21}
     p_{m}^{'}=\int_{{{\Omega }_{m}}}{{{\beta }_{FF}}\left( \theta  \right)}d\Omega 
 \end{equation}
where ${{\Omega }_{m}}$ is the solid angle along the scattered direction that can be seen by the receiver aperture of area ${{A}_{r}}$, and ${{\beta }_{FF}}\left( \theta  \right)$ is the FFPF defined earlier. The probability that is moves out of the receiver FOV is thus $\left( 1-p_{m}^{'} \right)$.

During the process, the photon moves with random step size, and along this path, we need to consider energy loss modeling. As the photon moves from the (m-1)th scatter or reflection center location ${{\mathbf{r}}_{m-1}}$ to the m-th scattering center location ${{\mathbf{r}}_{m}}$, the propagation length $\left| {{\mathbf{r}}_{m}}-{{\mathbf{r}}_{m-1}} \right|$ is given by the random variable in Eq.(\ref{eq14}). Moving along the scatter centers, the photon experiences an energy loss of $\exp \left( -c(\lambda )\left| {{\mathbf{r}}_{m}}-{{\mathbf{r}}_{m-1}} \right| \right)$. Impinging upon the m-th scattering center, the photon’s survival probability is degraded due to this energy loss and so is updated as\cite{30}:
\begin{equation}
\label{eq22}
     {{W}_{m}}=\left( 1-p_{m-1}^{'} \right)\exp \left( -c(\lambda )\left| {{\mathbf{r}}_{m}}-{{\mathbf{r}}_{m-1}} \right| \right){{W}_{m-1}}
\end{equation}

The photon successfully arrival probability after m-th scattering events as:
\begin{equation}
\label{eq23}
    {{P}_{m}}={{W}_{m}}p_{m}^{'}p_{m}^{''}
\end{equation}
where $p_{m}^{''}=\exp \left( -c(\lambda )\left| {{\mathbf{r}}_{m}}-{{\mathbf{r}}^{'}} \right| \right)$ represents the propagation loss of the photon from the m-th scattering center to the receiver, and ${{\mathbf{r}}^{'}}$ is the location vector of the receiver.

To obtain the CIR of the joint channel, it is essential to accord the propagation time of each photon, in addition to calculating the arrival probability of the photon. For a photon that reaches the receiver after m scattering interactions, the total migration path can be calculated as ${{d}_{m}}=\sum\nolimits_{i=1}^{m}{\left| {{\mathbf{r}}_{i}}-{{\mathbf{r}}_{i-1}} \right|}$. From this, we easily obtain the propagation time delay of a single photon as ${{\tau }_{m}}={{d}_{m}}/c$. We assume that a photon undergoes M scattering events, then corresponding to this photon is a set of arrival probabilities $\left( {{P}_{1}},{{P}_{2}},\ldots ,{{P}_{M}} \right)$and the corresponding propagation time delay$\left( {{\tau }_{1}},{{\tau }_{2}},\ldots ,{{\tau }_{M}} \right)$ between each scatter. By simulating the propagation behavior of all photons, we finally obtain a set of arrival times and the corresponding arrival probabilities and then normalize the arrival probabilities with the photon number to obtain the CIR.

 \begin{figure}[!htbp]
    \centering\includegraphics[width=6cm]{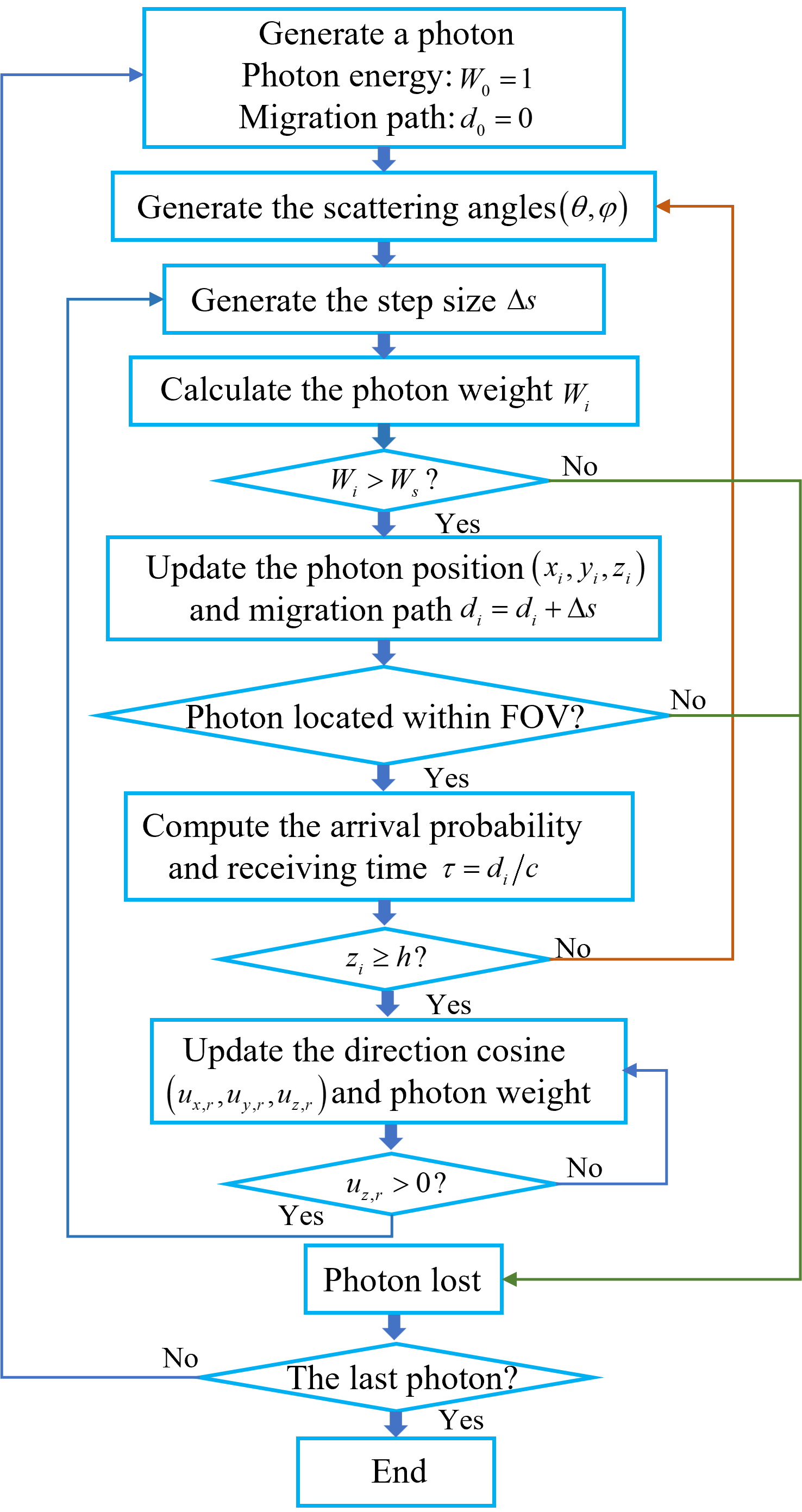}
    \caption{Algorithm structure for MCS process}
    \label{fig:4}
\end{figure}

We summarize the algorithm structure of our MCS process shown in Fig.\ref{fig:4}. Before implementing our simulation, we set the maximum number of simulation photons N(${{10}^{8}}$) and the weight threshold ${{W}_{s}}$ as the halt conditions. Each photon with an initial weight of 1 is emitted from the transmitter and its direction is determined by a pair of angles. Subsequent scattering occurs after the photon moves in that direction in random steps $\Delta s$. Then we compute the photon weight and determine whether it exceeds the set threshold ${{W}_{s}}$. If the weight is less than the ${{W}_{s}}$, the photon is considered lost. Otherwise, we update the photon position and migration path and then judge whether the photon is located within the FOV. If the photon is not located within the FOV, it is deemed lost. Conversely, we estimate its arrival probability and receiving time, and decide if it will be reflected or scattered based on its current position. Specifically, when the photon is above the sea surface, it is reflected by the rough surface. In contrast, it is scattered by the particle. For a reflection photon, we only recalculate its direction since we have ignored the position offset between reflection events as mentioned earlier. If the photon continues to move upward after reflection, it is reflected again. Otherwise, it undergoes the scattering process mentioned above. To obtain the final statistical result, we traverse all photons.

\subsection{Weighted Double Gamma Function Model}

In free space optical (FSO) communication links, several closed-form functions have been presented to model the impulse response. The single Gamma function is adopted by Geller to model the impulse response in clouds initially. Then the Gamma function has also been adopted in other scenarios such as atmosphere ultraviolet scattering channel and indoor optical wireless diffusion channel.

For underwater LOS links, Dong applied double Gamma function (DGF)first to model the impulse response for scattering channel in coastal and harbor water \cite{10}. The form of DGF can be written as:
\begin{equation}
\label{eq24}
    \hbar \left( t \right)={{C}_{1}}\Delta t{{e}^{-{{C}_{2}}\Delta t}}+{{C}_{3}}\Delta t{{e}^{-{{C}_{4}}\Delta t}}
\end{equation}
where ${{C}_{1}},{{C}_{2}},{{C}_{3}},{{C}_{4}}$ are the four parameters to be found; $\Delta t=t-{{t}_{0}}$, where $t$ is the time scale and ${{t}_{0}}$ is the earliest arrival time of photon detected by the receiver. 

However, the DGF is only applicable with relatively large values of the attenuation length. To generalize these functions, Dong added two parameters to the DGF and presented the weighted double gamma function (WDGF), written as \cite{31}:
\begin{equation}
\label{eq25}
    h(t)=\frac{{{C}_{1}}\cdot C_{2}^{-\alpha }}{\Gamma (\alpha )}\Delta {{t}^{\alpha -1}}{{e}^{-\Delta t/{{C}_{2}}}}+\frac{{{C}_{3}}\cdot C_{4}^{-\beta }}{\Gamma (\beta )}\Delta {{t}^{\beta -1}}{{e}^{-\Delta t/{{C}_{4}}}}
\end{equation}
here, $\Gamma \left( \cdot  \right)$is the Gamma function, $\alpha $ and $\beta $ are the two newly added parameters to be determined. The set of parameters $\left( {{C}_{1}},{{C}_{2}},{{C}_{3}},{{C}_{4}},\alpha ,\beta  \right)$ can be computed vis the Monte Carlo simulation with the nonlinear square criterion given by:
\begin{equation}
\label{eq26}
    \left( {{C}_{1}},{{C}_{2}},{{C}_{3}},{{C}_{4}},\alpha ,\beta  \right)=\arg \min \left( \int_{0}^{\infty }{{{\left[ {{h}_{MC}}\left( t \right)-{{h}_{DGF}}\left( t \right) \right]}^{2}}dt} \right)
\end{equation}
where ${{h}_{MC}}\left( t \right)$ is impulse response results obtained via MC simulation, ${{h}_{WDGF}}\left( t \right)$ is the WDGF model in Eq.(\ref{eq25}).

\section{Performance Evaluation}

In this section, we conduct the performance analysis of the UWOC NLOS system considering both scattering reflection and turbulence based on the close-formed WDGF model obtained in the previous section. We calculate the average bit error rate and outage probability performance for coastal and harbor water.

\subsection{BER Performance}

Considering intensity-modulated direct-detection (IM/DD) with on-off keying (OOK) modulation where the “ON” state signal will be transmitted with the rectangular pulse shape $P\left( t \right)$. The transmitted data sequence can be written as:
\begin{equation}
\label{eq27}
    S\left( t \right)=\sum\limits_{k=-\infty }^{\infty }{{{b}_{k}}}P\left( t-k{{T}_{b}} \right)
\end{equation}
where ${{b}_{k}}\in \left[ 0,1 \right]$is binary signal in the k-th time slot, ${{T}_{b}}=1/{{R}_{b}}$ is the single bit duration time, ${{R}_{b}}$ respects the data transmission rate and $P\left( t-k{{T}_{b}} \right)$is the optical signal with average power $2{{P}_{b}}$ in the k-th time slot. The received signal can be obtained by calculating the convolution of $S\left( t \right)$ and CIR $h\left( t \right)$:
\begin{equation}
\label{eq28}
    y\left( t \right)=S\left( t \right)*h\left( t \right)=\tilde{h}\sum\limits_{k=-\infty }^{\infty }{{{b}_{k}}\Gamma \left( t-k{{T}_{b}} \right)+n\left( t \right)}
\end{equation}
where $h\left( t \right)=\tilde{h}{{h}_{0}}\left( t \right)$represents the channel aggregated impulse response, $\Gamma \left( t \right)=P\left( t \right)*{{h}_{0}}\left( t \right)$, * respects the convolution operator, $n\left( t \right)$ represents the additive noise component. 

Based on the Eq.(\ref{eq28}), the ${{0}^{th}}$ time slot integrated current can be evaluated by:
\begin{equation}
\label{eq29}
    {{r}_{{{b}_{0}}}}={{b}_{0}}\tilde{h}{{u}^{\left( I,k=0 \right)}}+\tilde{h}\sum\limits_{k=-L}^{-1}{{{b}_{k}}}{{u}^{\left( I,k \right)}}+{{\upsilon }_{n}}
\end{equation}
where ${{u}^{\left( I,0 \right)}}$ respects the contribution of the desired signal with the expression ${{u}^{\left( I,0 \right)}}=\Re\int_{0}^{{{T}_{b}}}{\Gamma \left( t \right)}dt$, while ${{u}^{\left( I,k\ne 0 \right)}}$ refers to the inter-symbol interference (ISI) effects, written as ${{u}^{\left( I,k \right)}}=\Re\int_{-k{{T}_{b}}}^{-\left( k-1 \right){{T}_{b}}}{\Gamma \left( t \right)}dt$. Here $\Re$ is the photodetector’s responsivity. Additional term ${{\upsilon }_{n}}$ indicates the integration of the receiver Gaussian-distribution noise component with mean zero and variance $\sigma _{{{T}_{b}}}^{2}$.

Assuming the channel information is available to the receiver, the information transmission error probability can be obtained by integrating the current over symbol interval time ${{T}_{b}}$ and comparing it with the threshold value of $\tilde{h}{{u}^{\left( I,0 \right)}}/2$. Therefore, the probability of error on transmission of symbols “1” and “0” can be obtained respectively as:
\begin{equation}
\label{eq30}
    \begin{aligned}
  & {{P}_{be|1,\tilde{h},{{b}_{k}}}}=\Pr \left( {{r}_{{{b}_{0}}}}\le \tilde{h}{{u}^{\left( I,0 \right)}}/2|{{b}_{k}}=1 \right) \\ 
 & \quad \quad \,=Q\left( \frac{\tilde{h}\left[ {{u}^{\left( I,0 \right)}}/2+\sum\nolimits_{k=-L}^{-1}{{{b}_{k}}{{u}^{\left( I,k \right)}}} \right]}{{{\sigma }_{{{T}_{b}}}}} \right), \\ 
 & {{P}_{be|0,\tilde{h},{{b}_{k}}}}=\Pr \left( {{r}_{{{b}_{0}}}}\ge \tilde{h}{{u}^{\left( I,0 \right)}}/2|{{b}_{k}}=0 \right) \\ 
 & \quad \quad \,=Q\left( \frac{\tilde{h}\left[ {{u}^{\left( I,0 \right)}}/2-\sum\nolimits_{k=-L}^{-1}{{{b}_{k}}{{u}^{\left( I,k \right)}}} \right]}{{{\sigma }_{{{T}_{b}}}}} \right), \\ 
\end{aligned}
\end{equation}
where $Q\left( x \right)=\left( 1/\sqrt{2\pi } \right)\int_{x}^{\infty }{\exp \left( -{{y}^{2}}/2 \right)}dy$ is the Gaussian-Q function. Then the average BER can be obtained by average all ${{2}^{L}}$ possible data sequences for ${{b}_{k}}$’s as:
\begin{equation}
\label{eq31}
    {{P}_{be}}=\frac{1}{{{2}^{L+1}}}\sum\limits_{{{b}_{k}}}{\int_{0}^{\infty }{\left[ {{P}_{e|1,{{b}_{k}}}}+{{P}_{e|0,{{b}_{k}}}} \right]}}{{f}_{{\tilde{h}}}}\left( {\tilde{h}} \right)d\tilde{h}.
\end{equation}

Observing Eq.(\ref{eq31}), it can be found that the interior of the integral is the sum of the Gaussian Q-functions of two different independent variables, and to represent the BER of different symbols in a uniform way, it is defined that ${{P}_{be|{{b}_{0}},{{b}_{k}}}}$ denotes the average BER corresponding to ${{b}_{0}}=1$ and 0 in Eq.(\ref{eq30}):
\begin{equation}
\label{eq32}
    \begin{aligned}
  & {{P}_{be|{{b}_{0}},{{b}_{k}}}}=\int_{0}^{\infty }{{{P}_{be|{{b}_{0}},\tilde{h},{{b}_{k}}}}}{{f}_{{\tilde{h}}}}\left( {\tilde{h}} \right)d\tilde{h} \\ 
 & \quad \quad \ \ \,=\int_{0}^{\infty }{Q\left( {{C}_{{{b}_{0}}}}\tilde{h} \right)}{{f}_{{\tilde{h}}}}\left( {\tilde{h}} \right)d\tilde{h} \\ 
\end{aligned}
\end{equation}
where ${{C}_{{{b}_{0}}}}$ is the constant coefficient of the Gaussian Q-function in Eq.(\ref{eq32}) with the expression as:
\begin{equation}
\label{eq33}
    {{C}_{{{b}_{0}}}}=\frac{\left[ {{u}^{\left( I,0 \right)}}/2+{{(-1)}^{{{b}_{0}}+1}}\sum\nolimits_{k=-L}^{-1}{{{b}_{k}}{{u}^{\left( I,k \right)}}} \right]}{{{\sigma }_{{{T}_{b}}}}}
\end{equation}

Then the closed-form expression for the average BER can be obtained as:
\begin{equation}
\label{eq34}
    {{P}_{be}}=\frac{1}{{{2}^{L+1}}}\sum\nolimits_{{{b}_{k}}}{\left[ {{P}_{be|1,{{b}_{k}}}}+{{P}_{be|0,{{b}_{k}}}} \right]}
\end{equation}

\subsection{Outage Probability}

The outage probability is defined as the probability that the channel capacity falls below a predetermined rate target, and corresponds to the probability that the instantaneous signal-to-noise ratio $\gamma $ at the receiver falls below a certain threshold ${{\gamma }_{th}}$:
\begin{equation}
\label{eq35}
    {{P}_{out}}=\Pr \left( \gamma <{{\gamma }_{th}} \right)
\end{equation}

The prerequisite for calculating the outage probability is to define the signal-to-noise ratio formula for the channel. It is defined as the ratio of the effective signal to the sum of the interference and Gaussian white noise, i.e:
\begin{equation}
\label{eq36}
    \gamma =\frac{{{{\tilde{h}}}^{2}}{{\left[ {{u}^{(I,0)}}+{{(-1)}^{{{b}_{0}}+1}}\sum\nolimits_{k=-L}^{-1}{2{{b}_{k}}{{u}^{(I,k)}}} \right]}^{2}}}{\sigma _{{{T}_{b}}}^{2}}
\end{equation}

Based on this defining equation, the outage probability of the UWOC system is expressed as:
\begin{equation}
\label{eq37}
    {{P}_{out}}=\frac{1}{{{2}^{L}}}\sum\nolimits_{\left\{ {{b}_{k}} \right\}}{\Pr \left( \tilde{h}<\Upsilon  \right)}=\frac{1}{{{2}^{L}}}\sum\nolimits_{\left\{ {{b}_{k}} \right\}}{{{F}_{{\tilde{h}}}}}\left( \Upsilon  \right)
\end{equation}
where ${{F}_{{\tilde{h}}}}\left( \centerdot  \right)$ denotes the cumulative distribution function of the fading coefficient $\tilde{h}$. For easy calculation, the constant term is denoted as:
\begin{equation}
\label{eq38}
    \Upsilon =\sqrt{\frac{{{\gamma }_{th}}\sigma _{{{T}_{b}}}^{2}}{{{\left[ {{u}^{(I,0)}} \right]}^{2}}-{{\gamma }_{th}}{{\left[ \sum\nolimits_{k=-L}^{-1}{{{b}_{k}}{{u}^{(I,k)}}} \right]}^{2}}}}
\end{equation}

Eq.(\ref{eq37}) shows that the outage probability requires an integration operation over the probability density function, and for channels with a lognormal distribution, the cumulative distribution function is calculated as follows:
\begin{equation}
\label{eq39}
    {{F}_{{\tilde{h}}}}\left( \Upsilon  \right)=1-Q\left( \frac{\ln \left( \Upsilon  \right)+2\sigma _{X}^{2}}{2{{\sigma }_{X}}} \right)
\end{equation}

The outage probability can be obtained by bringing the above equation into Eq.(\ref{eq37}).

\begin{figure*}[!ht]
\centering
\subfigure[]{\label{fig5:subfig:a}
\includegraphics[width=0.45\linewidth]{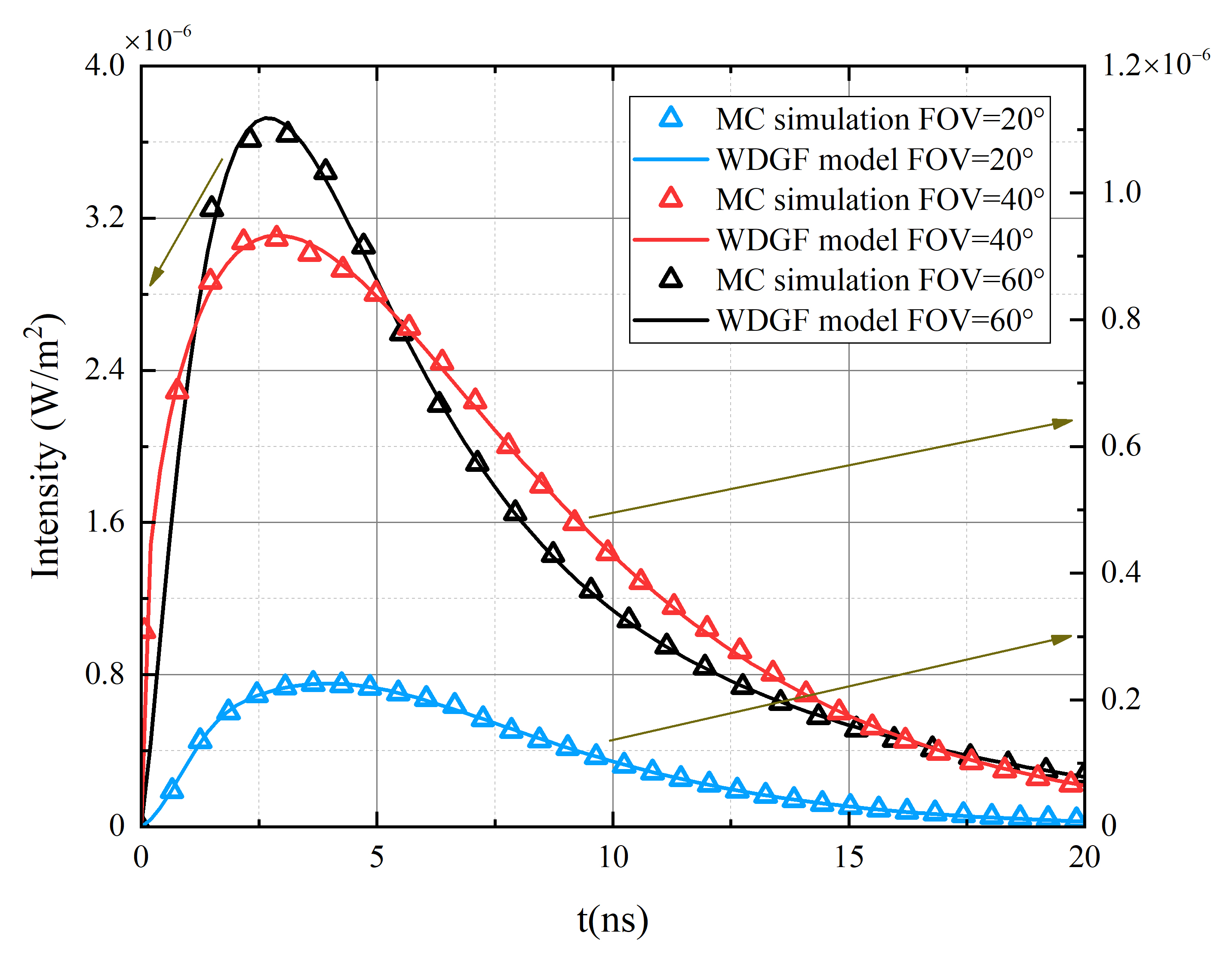}}
\hspace{0.01\linewidth}
\subfigure[]{\label{fig5:subfig:b}
\includegraphics[width=0.45\linewidth]{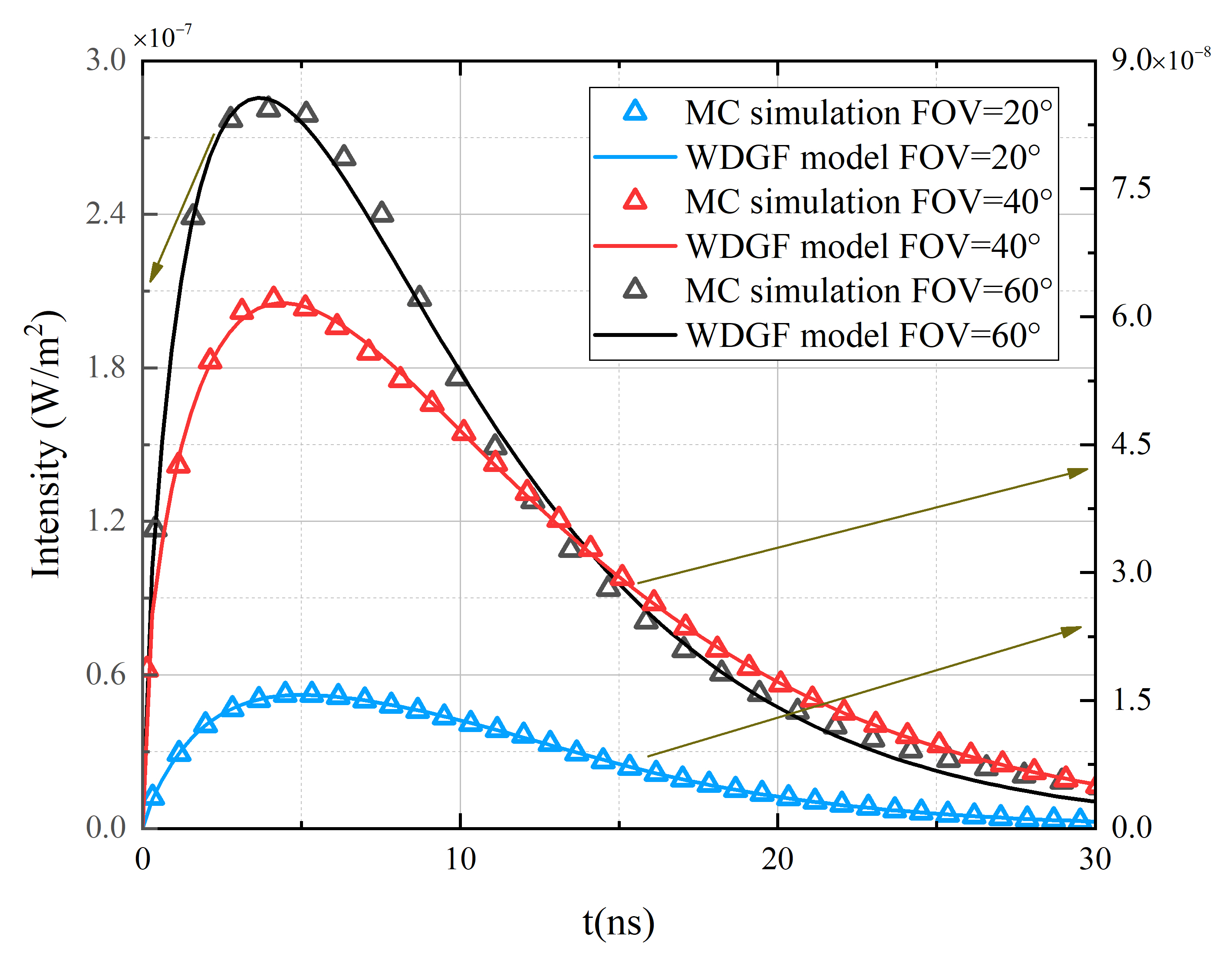}}
\vfill
\subfigure[]{\label{fig5:subfig:c}
\includegraphics[width=0.45\linewidth]{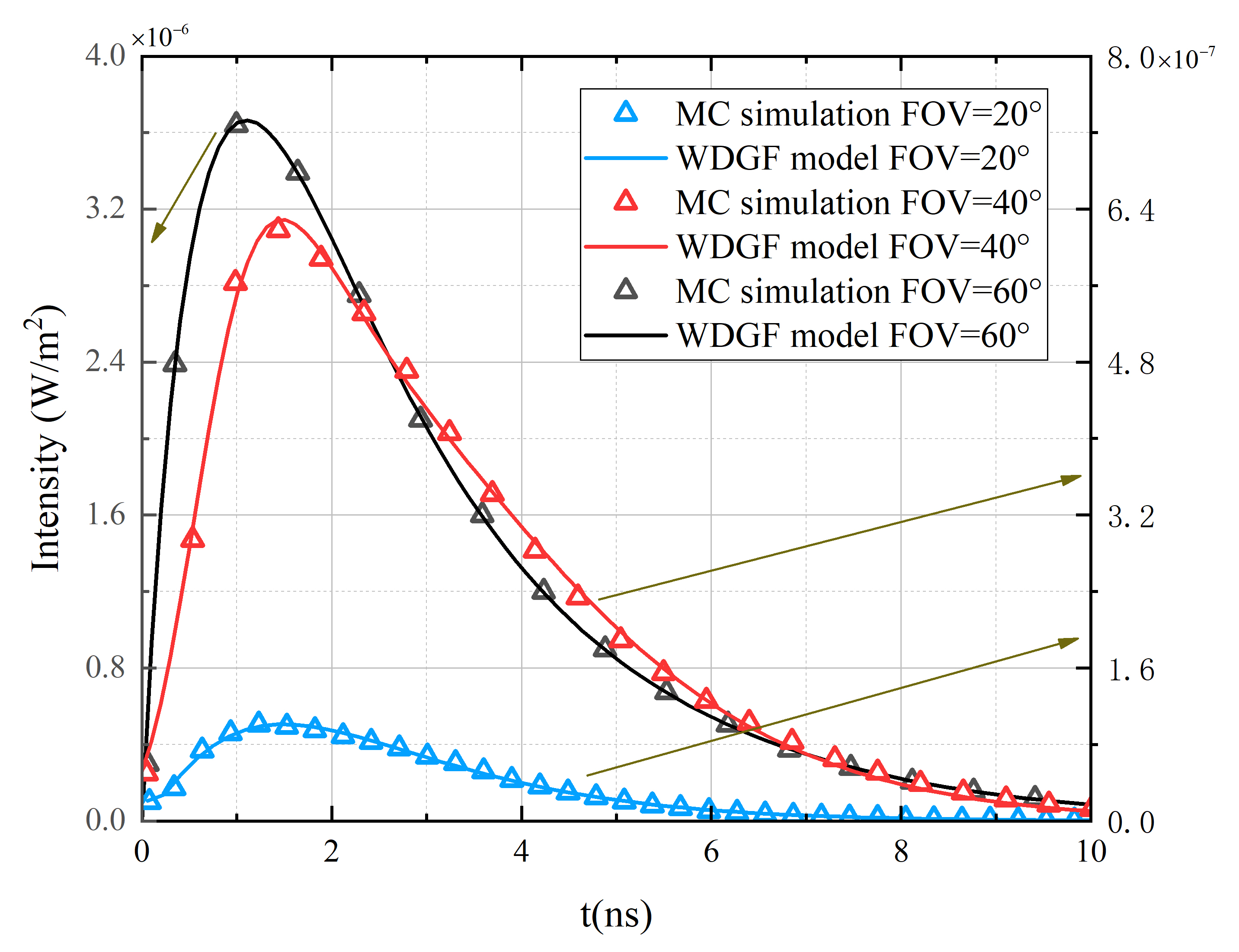}}
\hspace{0.01\linewidth}
\subfigure[]{\label{fig5:subfig:d}
\includegraphics[width=0.45\linewidth]{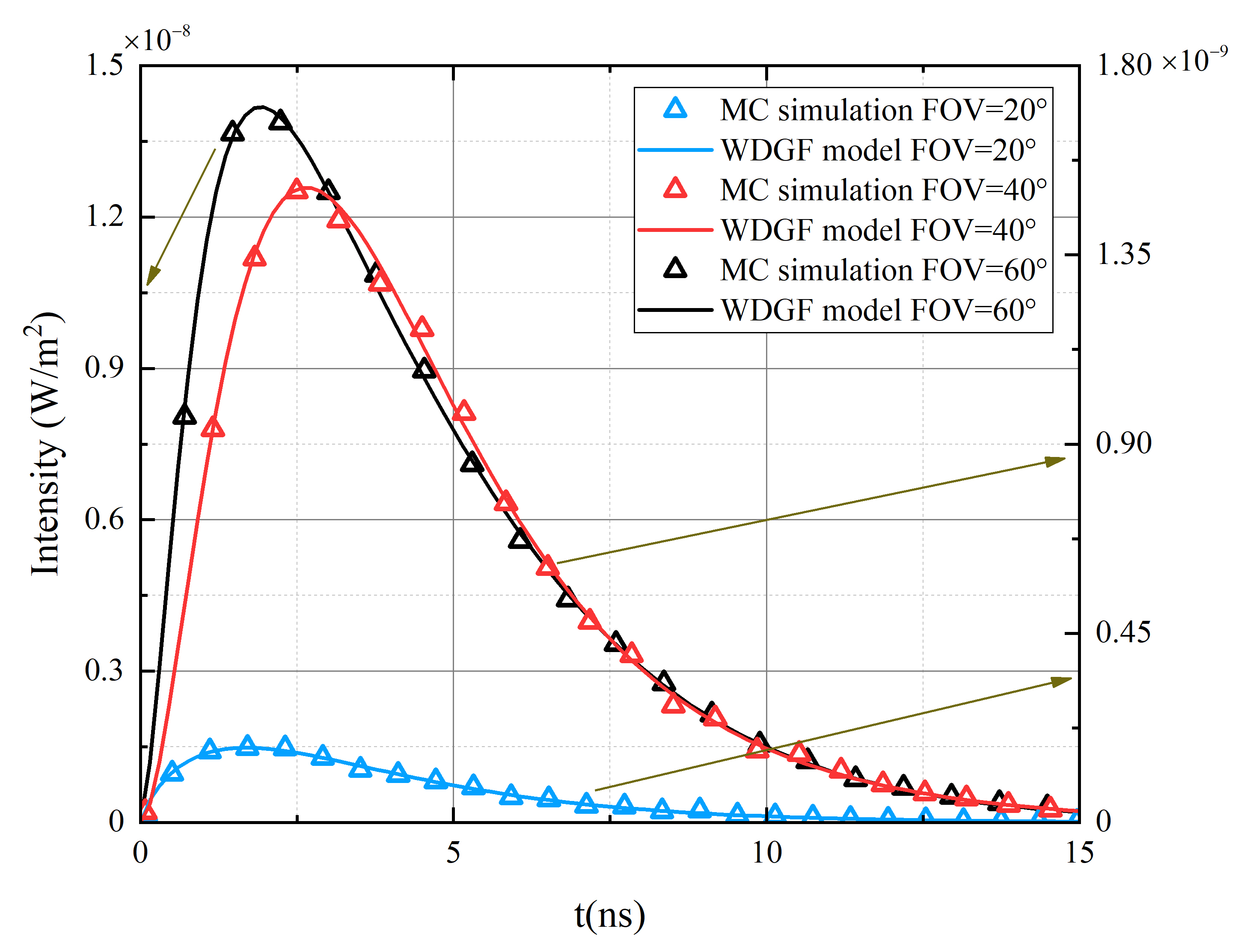}}
\centering
\caption{Impulse response in costal and harbor water. (a) L=10m, coastal water (b) L=20m, coastal water. (c) L=5m, harbor water. (d) L=10m, harbor water.}
\label{fig:5}
\end{figure*}

\section{Numerical results}

In this section, we provide the numerical results for the MCS-based CIR model that takes into account the effects of reflection based on rough surface and seawater scattering. We consider a UWOC system with a 532 nm wavelength light source and a photon detector with an aperture of 50 cm. ${{10}^{8}}$ photons are simulated to obtain the CIR using MATLAB in coastal and harbor water. Moreover, the Average BER and channel bandwidth of the UWOC NLOS link are analyzed based on the results of WDGF obtained from MCS fitting.

\subsection{Impulse Response}
\begin{table}[!h]
\centering
\caption{\bf Parameters of WDGF in Different channels}
\begin{tabular}{ccccccc}
\hline
FOV & ${{C}_{1}}$ & ${{C}_{1}}$ & ${{C}_{1}}$ & ${{C}_{1}}$ & $\alpha $ & $\beta $ \\
\hline
\multicolumn{3}{c}{Coastal water, L=10m       } \\
$20{}^\circ $ & $2.04e-6$ & $2.85$ & $4.85e-6$ & $0.31$ & $2.44$ & $6.21$ \\
$40{}^\circ $ & $3.57e-6$ & $4.05$ & $5.99e-6$ & $3.90$ & $1.28$ & $2.11$ \\
$60{}^\circ $ & $1.09e-5$ & $5.01$ & $1.15e-5$ & $1.78$ & $2.16$ & $2.33$ \\
\multicolumn{3}{c}{Coastal water, L=20m       } \\
$20{}^\circ $ & $1.45e-8$ & $1.47$ & $2.19e-7$ & $5.58$ & $8.75$ & $1.85$ \\
$40{}^\circ $ & $9.37e-7$ & $6.71$ & $e-8$ & $3.90$ & $1.28$ & $2.11$ \\
$60{}^\circ $ & $1.09e-5$ & $5.01$ & $1.15e-5$ & $1.78$ & $2.16$ & $2.33$ \\
\multicolumn{3}{c}{Harbor water, L=5m       } \\
$20{}^\circ $ & $3.48e-7$ & $1.12$ & $1.51e-8$ & $0.02$ & $2.32$ & $0.11$ \\
$40{}^\circ $ & $1.46e-8$ & $0.01$ & $2.40e-6$ & $2.01$ & $1.2e-5$ & $1.39$ \\
$60{}^\circ $ & $1.07e-5$ & $1.12$ & $2.43e-6$ & $1.31$ & $1.94$ & $4.17$ \\
\multicolumn{3}{c}{Harbor water, L=20m       } \\
$20{}^\circ $ & $7.61e-9$ & $2.02$ & $1.72e-9$ & $2.45$ & $1.79$ & $1.88$ \\
$40{}^\circ $ & $1.15e-8$ & $15.39$ & $7.58e-8$ & $2.07$ & $0.10$ & $1.81$ \\
$60{}^\circ $ & $6.51e-7$ & $1.98$ & $1.11e-7$ & $0.64$ & $2.30$ & $3.18$ \\
\hline
\end{tabular}
  \label{tab:2}
\end{table}

Fig.\ref{fig:5} shows a CIR and WDGF fitting for different FOV values and we set the transceiver baseline distance $L$ to be 10m, 20m for coastal water and 5m, 10m for harbor water respectively. The parameters of the WDGF fitting are listed in Table. \ref{tab:2}. Note that the start time of CIR is shifted from ${{t}_{0}}$ to 0. The temporal dispersion can be evaluated from the WDGF model, it is defined as the time interval of CIR falling 20 dB below the peak. The temporal dispersion for different water types and link ranges are shown in Tab. \ref{tab:3}. By comparing the results, we can draw several conclusions. First, the WDGF fits well with the simulation results regardless of water type, FOV, and propagation distance. Second, with the increase in distance, the amplitude of the CIR decreases obviously, and the time delay rises significantly. The reason is that the photons undergo more scattering and attenuation over a longer distance. Third, the temporal pulse spread and CIR amplitude increase with an increasing FOV. This is explained by the fact that a larger FOV can detect more photons, thus increasing the received power. In addition, we find the pulse spread is smaller in harbor water compared to coastal water. The basic reason is that the high attenuation coefficient of harbor water leads to a sharp decrease in the number of photons reaching the receiver, which reduces the received power and pulse spread.

\begin{table}[htbp]
    \centering
    \caption{\bf Temporal dispersion for various ranges, water types and FOVs.}
\begin{tabular}{cccccc}
\hline
\multicolumn{3}{c}{Coastal water} & \multicolumn{3}{c}{Harbor water} \\
\hline
FOV & L=10m & L=20m & FOV & L=5m & L=10m \\
$20{}^\circ $ & $24.56ns$ & $40.23ns$ &  $20{}^\circ $ & $8.89ns$ & $13.84ns$ \\
$40{}^\circ $ & $29.01ns$ & $45.27ns$ &  $40{}^\circ $ & $10.61ns$ & $14.71ns$ \\
$60{}^\circ $ & $31.98ns$ & $47.58ns$ &  $60{}^\circ $ & $11.85ns$ & $15.79ns$ \\
    \hline
    \end{tabular}
    \label{tab:3}
\end{table}

We introduce the root mean square errors (RMSE) to compare the simulation results of CIR with the WDGF fitting results. The RMSE is summarized in Tab. \ref{tab:4}. The smaller the root mean square error illustrates a better fitting performance. We can verify that the RMSE for each scenario is always less than $0.05$. Therefore, we can conclude that the WDGF can model the CIR of UWOC NLOS links both considering the scattering and reflection effects caused by particle and rough surface respectively.
 \begin{table}[htbp]
    \centering
    \caption{\bf RMSE value of WDGF fitting curve for various channels.}
\begin{tabular}{cccccc}
\hline
\multicolumn{3}{c}{Coastal water} & \multicolumn{3}{c}{Harbor water} \\
\hline
FOV & L=10m & L=20m & FOV & L=5m & L=10m \\
$20{}^\circ $ & $0.02171$ & $0.04068$ &  $20{}^\circ $ & $0.01523$ & $0.01198$ \\
$40{}^\circ $ & $0.00535$ & $0.01928$ &  $40{}^\circ $ & $0.01264$ & $0.07029$ \\
$60{}^\circ $ & $0.00874$ & $0.04724$ &  $60{}^\circ $ & $0.00284$ & $0.02993$ \\
    \hline
    \end{tabular}
    \label{tab:4}
\end{table}
\begin{figure*}[!htbp]
\centering
\subfigure[]{
\label{fig6:subfig:a} %% label for first subfigure
\includegraphics[width=0.48\linewidth]{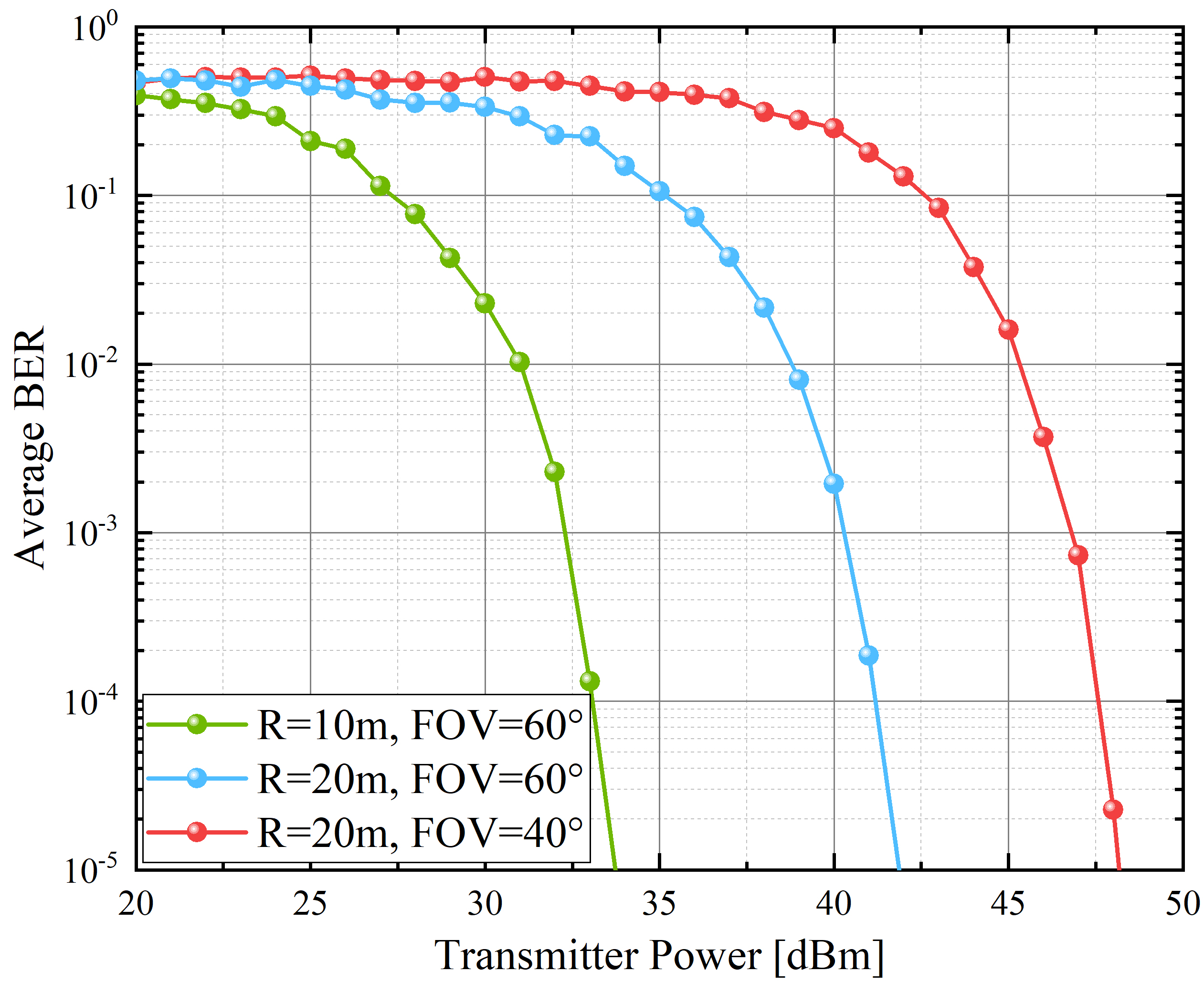}}
\hspace{0.02in}
\subfigure[]{
\label{fig6:subfig:b} %% label for second subfigure
\includegraphics[width=0.48\linewidth]{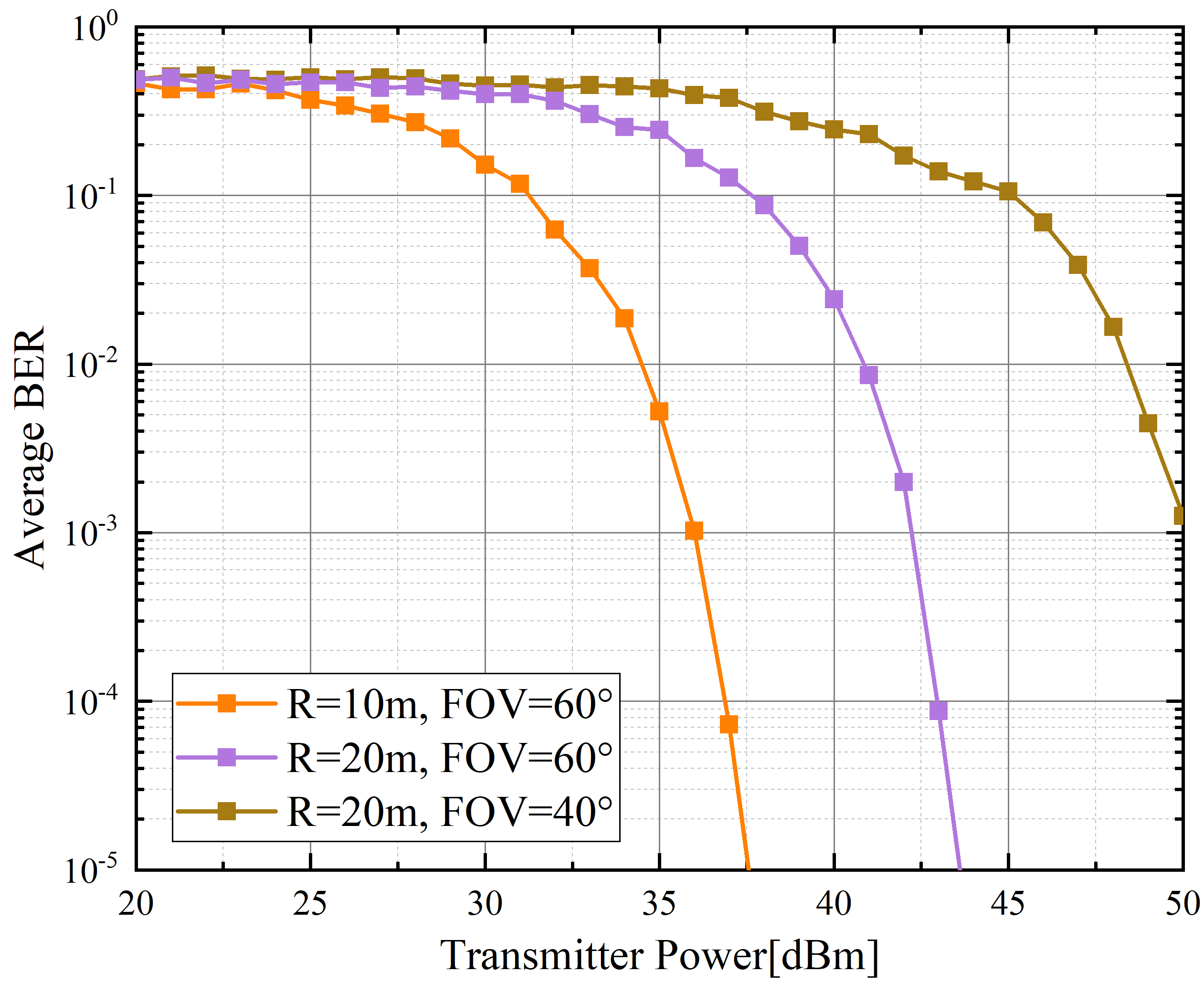}}
\centering
\caption{ BER performance for different link ranges and FOVs under coastal water with two scintillation index:(a) $\sigma _{I}^{2}=0.2$ (b) $\sigma _{I}^{2}=0.8$.}
\label{fig:6} %% label for entire figure
\end{figure*}

\subsection{Performance analysis}
\begin{figure*}[!h]
\centering
\subfigure[]{
\label{fig7:subfig:a} %% label for first subfigure
\includegraphics[width=0.48\linewidth]{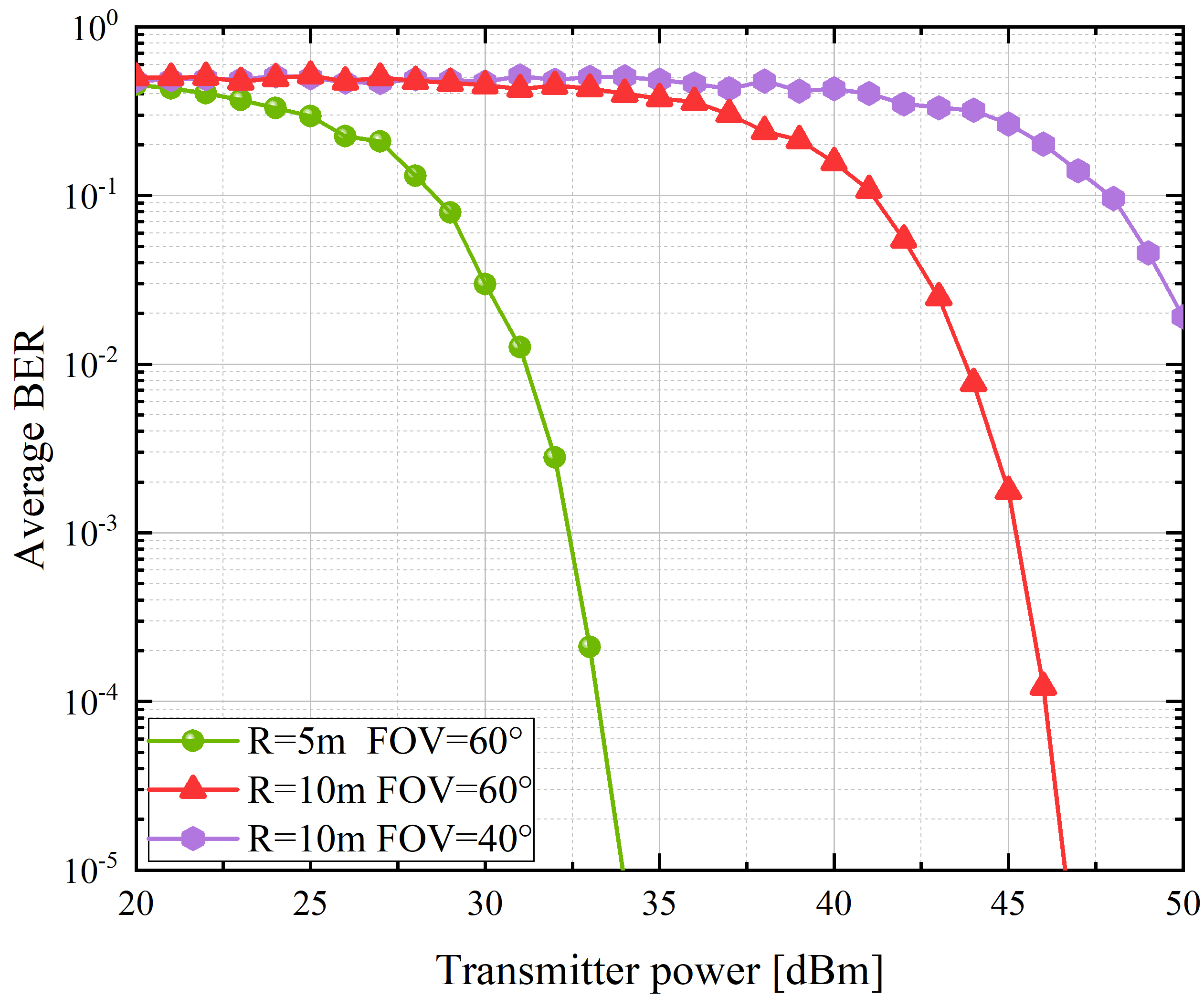}}
\hspace{0.02in}
\subfigure[]{
\label{fig7:subfig:b} %% label for second subfigure
\includegraphics[width=0.48\linewidth]{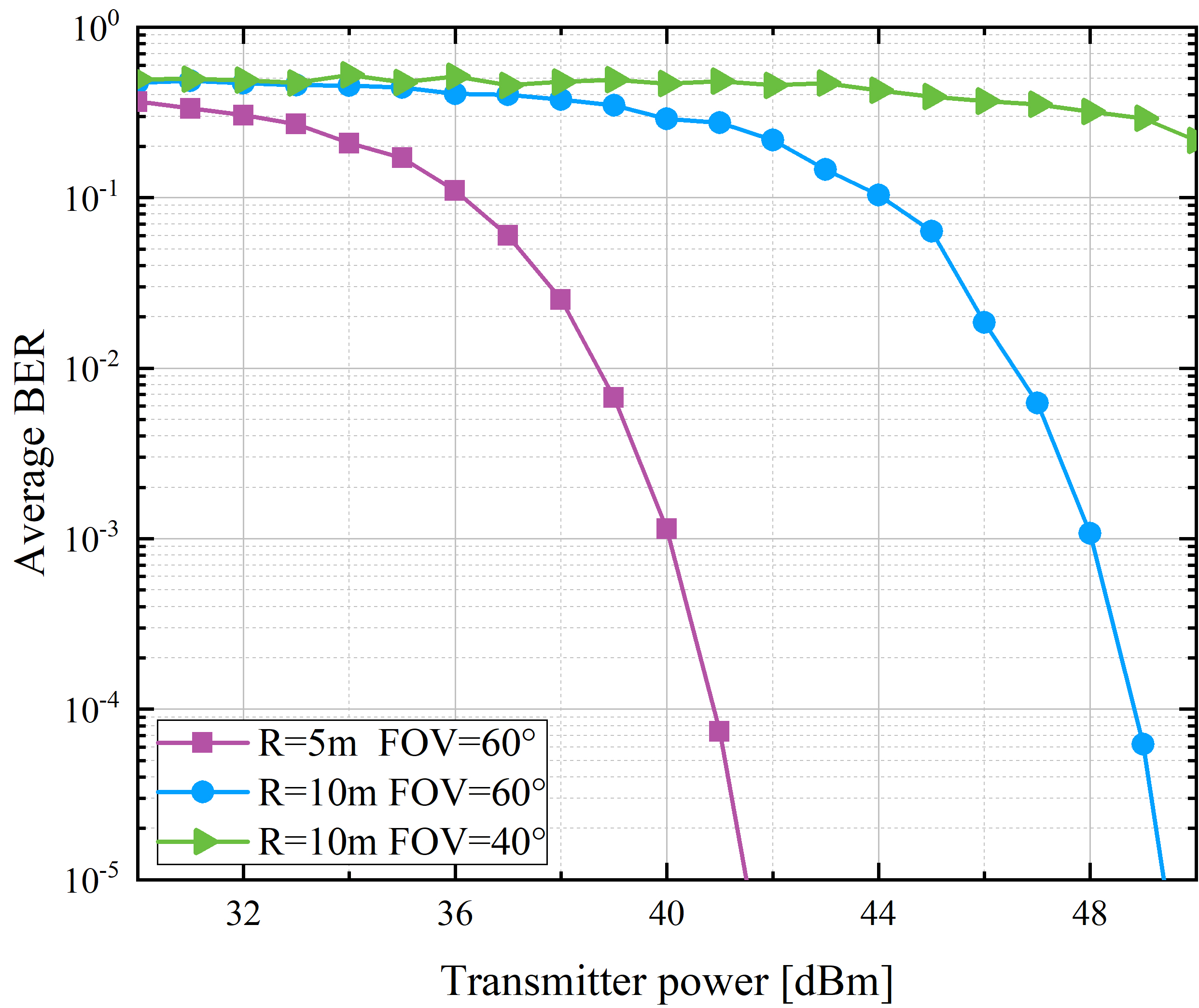}}
\centering
\caption{ BER performance for different link ranges and FOVs under harbor water with two scintillation index:(a) $\sigma _{I}^{2}=0.2$ (b) $\sigma _{I}^{2}=0.8$.}
\label{fig:7} %% label for entire figure
\end{figure*}

Fig.\ref{fig:6} illustrates the BER for different turbulence conditions and channel parameters. Comparing Fig.\ref{fig6:subfig:a} and \ref{fig6:subfig:b} it can be observed that larger scintillation coefficients correspond to worse BER performance when the transceiver spacing and FOV angle are constant. When the turbulence conditions are constant, a larger FOV angle results in better BER performance, which is because a larger FOV angle increases the probability of a photon reaching the receiver. In addition, for the same FOV angle, the larger the distance between the transceiver and the receiver, the larger the BER, which is due to the greater attenuation that accompanies longer distances.

Fig.\ref{fig:7} describes the average BER values for harbor water under different turbulence conditions. The same conclusion as in Fig.\ref{fig:6} can be drawn by observing the BER curves: the larger the turbulence scintillation coefficient, the worse the BER performance. In addition, comparing different seawater conditions, it can be found that the BER performance is worse in turbid seawater, which is because the light in turbid water is scattered more times and has a lower probability of reaching the receiver. The combined simulation results show that a larger receiver FOV is favorable for obtaining better BER performance.

\begin{figure*}[!h]
\centering
\subfigure[]{
\label{fig8:subfig:a} %% label for first subfigure
\includegraphics[width=0.48\linewidth]{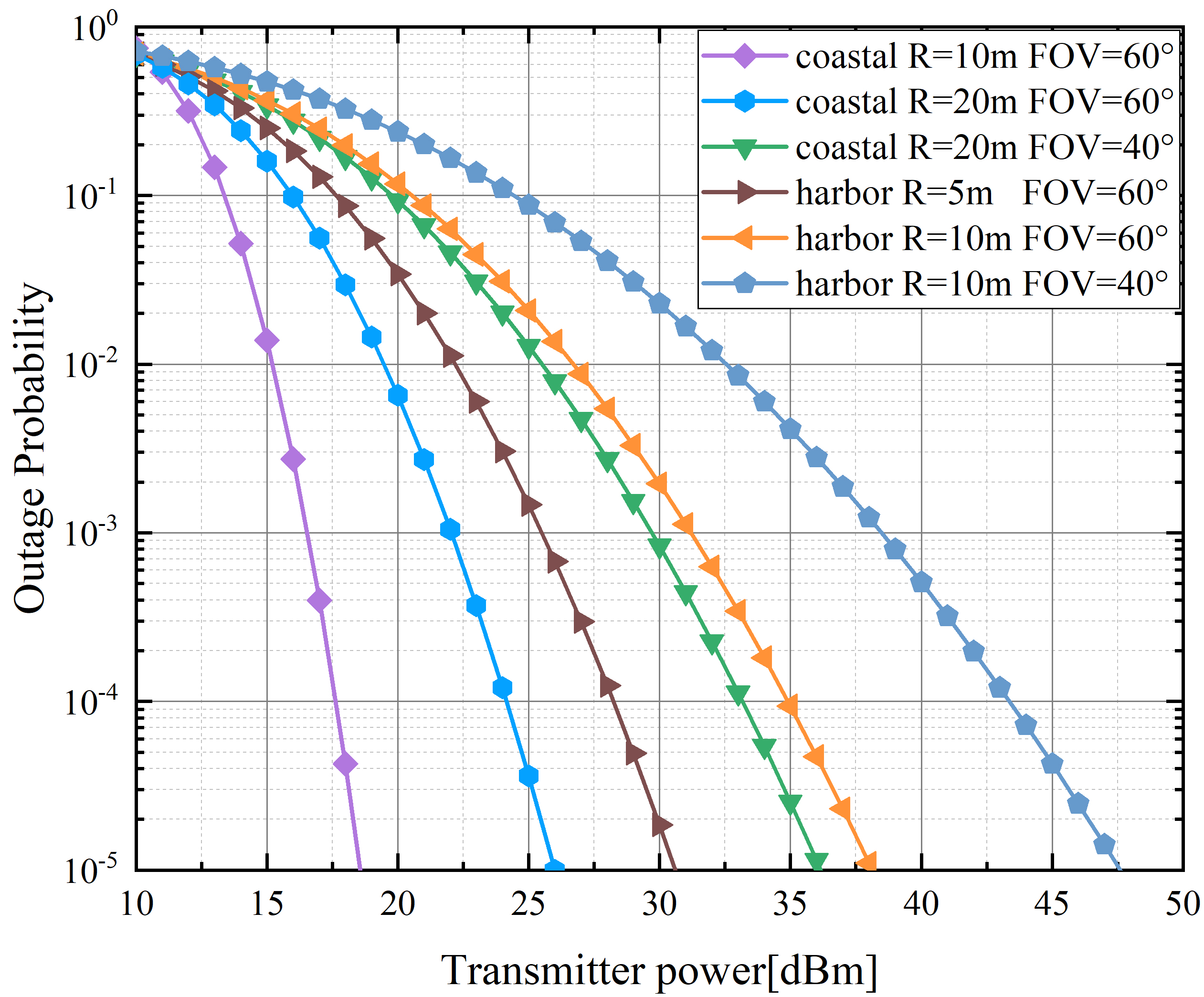}}
\hspace{0.02in}
\subfigure[]{
\label{fig8:subfig:b} %% label for second subfigure
\includegraphics[width=0.48\linewidth]{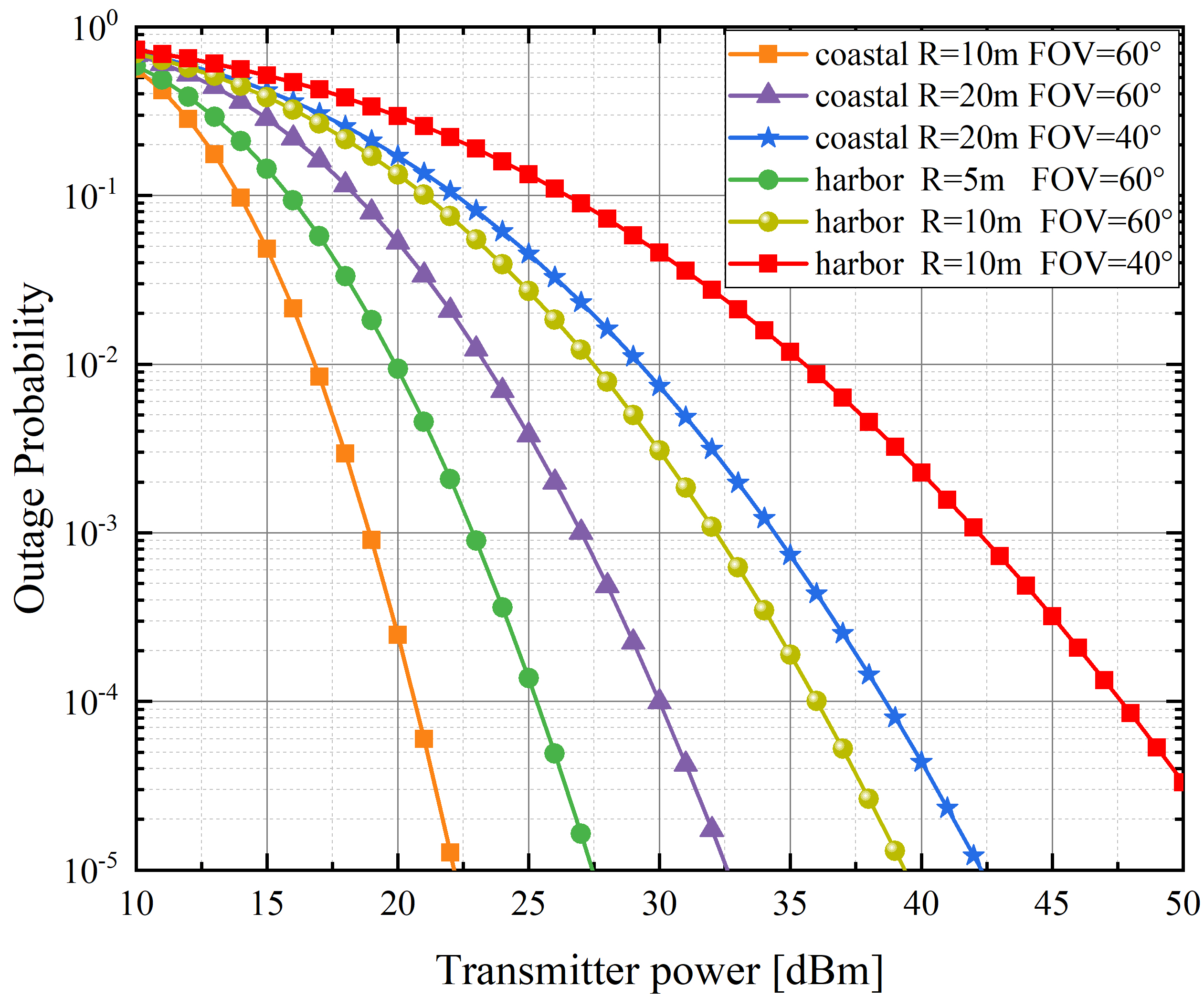}}
\centering
\caption{ Outage probability of different channel condition with: (a) $\sigma _{I}^{2}=0.2$ (b) $\sigma _{I}^{2}=0.8$.}
\label{fig:8} %% label for entire figure
\end{figure*}

Fig.\ref{fig:8} illustrates the outage probabilities for different channel and turbulence parameters, with plots a and b corresponding to the $\sigma _{I}^{2}=0.2$ and $\sigma _{I}^{2}=0.8$ conditions, respectively. Comparing the two plots it can be observed that larger turbulence scintillation coefficients correspond to higher outage probabilities. In addition, the performance of the interruption probability for the turbid water condition is worse than that of the coastal seawater. Larger FOV angles correspond to better interruption probability performance, which is the same conclusion as in Fig.\ref{fig:6} and \ref{fig:7}.

\section{Conclusion}
During underwater wireless NLOS channel modeling, photons may reach the receiver through seawater scattering or sea surface reflection. In this paper, we consider the above two factors and construct a channel model under turbid seawater conditions with the help of the MCS method and fit the CIR simulation results with a WDGF. Based on the CIR model, we obtain the channel time spreading information. Based on this closed-form expression, we analyze the BER and outage probability performance of the system in the presence of turbulent conditions. The results show that the communication performance is worse at larger turbulence scintillation coefficients and that larger beam dispersion angles are beneficial to compensate for the performance difference. The work done in this paper provides theoretical guidance for the construction of underwater wireless optical communication systems.

\subsection* {Funding.}
Fundamental Research Funds for the Central Universities
(JB160110, XJS16051); 111 Project (B08038); National Natural Science Foundation of China (61505155, 61571367).

\subsection*{Disclosures.}
The authors declare no conflicts of interest

\subsection*{Data availability.}
Data underlying the results presented in this paper are not publicly available at this time but may be obtained from the authors upon reasonable request.
%%%%%%%%%%%%%%%%%%%%%%% References %%%%%%%%%%%%%%%%%%%%%%%%%

%%%%%%%%%% If using BibTeX:
\bibliography{sample}

\begin{thebibliography}{10}
\newcommand{\enquote}[1]{``#1''}

\bibitem{1}
X.~Sun, C.~H. Kang, M.~Kong, \emph{et~al.}, \enquote{A review on practical considerations and solutions in underwater wireless optical communication,} {\protect\JournalTitle{Journal of Lightwave Technology}} \textbf{38}, 421--431 (2020).

\bibitem{2}
P.~Sajmath, R.~V. Ravi, and K.~A. Majeed, \enquote{Underwater wireless optical communication systems: A survey,} in \emph{2020 7th International Conference on Smart Structures and Systems (ICSSS),}  (IEEE, 2020), pp. 1--7.

\bibitem{3}
A.~S. Mohammed, S.~A. Adnan, M.~A.~A. Ali, and W.~K. Al-Azzawi, \enquote{Underwater wireless optical communications links: Perspectives, challenges and recent trends,} {\protect\JournalTitle{Journal of Optical Communications}}  (2022).

\bibitem{new1}
Z.~Jia, W.~Cheng, and H.~Zhang, \enquote{A partial learning-based detection scheme for massive mimo,} {\protect\JournalTitle{IEEE Wireless Communications Letters}} \textbf{8}, 1137--1140 (2019).

\bibitem{new2}
W.~Cheng, X.~Zhang, and H.~Zhang, \enquote{Optimal power allocation with statistical qos provisioning for d2d and cellular communications over underlaying wireless networks,} {\protect\JournalTitle{IEEE Journal on Selected Areas in Communications}} \textbf{34}, 151--162 (2016).

\bibitem{new3}
J.~Wang and W.~Cheng, \enquote{Heterogeneous quality of experience guarantees over wireless networks,} {\protect\JournalTitle{China Communications}} \textbf{15}, 51--59 (2018).

\bibitem{4}
C.~Li, K.-H. Park, and M.-S. Alouini, \enquote{On the use of a direct radiative transfer equation solver for path loss calculation in underwater optical wireless channels,} {\protect\JournalTitle{IEEE Wireless Communications Letters}} \textbf{4}, 561--564 (2015).

\bibitem{5}
S.~Jaruwatanadilok, \enquote{Underwater wireless optical communication channel modeling and performance evaluation using vector radiative transfer theory,} {\protect\JournalTitle{IEEE Journal on Selected Areas in Communications}} \textbf{26}, 1620--1627 (2008).

\bibitem{6}
C.~Gabriel, M.-A. Khalighi, S.~Bourennane, \emph{et~al.}, \enquote{Channel modeling for underwater optical communication,} in \emph{2011 IEEE GLOBECOM Workshops (GC Wkshps),}  (2011), pp. 833--837.

\bibitem{7}
C.~Gabriel, M.-A. Khalighi, S.~Bourennane, \emph{et~al.}, \enquote{Monte-carlo-based channel characterization for underwater optical communication systems,} {\protect\JournalTitle{J. Opt. Commun. Netw.}} \textbf{5}, 1--12 (2013).

\bibitem{8}
J.~Li, Y.~Ma, Q.~Zhou, \emph{et~al.}, \enquote{{Monte Carlo study on pulse response of underwater optical channel},} {\protect\JournalTitle{Optical Engineering}} \textbf{51}, 066001 (2012).

\bibitem{9}
S.~Tang, X.~Zhang, and Y.~Dong, \enquote{On impulse response for underwater wireless optical links,} in \emph{2013 MTS/IEEE OCEANS - Bergen,}  (2013), pp. 1--4.

\bibitem{10}
S.~Tang, Y.~Dong, and X.~Zhang, \enquote{Impulse response modeling for underwater wireless optical communication links,} {\protect\JournalTitle{IEEE Transactions on Communications}} \textbf{62}, 226--234 (2014).

\bibitem{11}
Y.~Li, M.~S. Leeson, and X.~Li, \enquote{Impulse response modeling for underwater optical wireless channels,} {\protect\JournalTitle{Appl. Opt.}} \textbf{57}, 4815--4823 (2018).

\bibitem{12}
R.~Boluda-Ruiz, P.~Rico-Pinazo, B.~Castillo-Vázquez, \emph{et~al.}, \enquote{Impulse response modeling of underwater optical scattering channels for wireless communication,} {\protect\JournalTitle{IEEE Photonics Journal}} \textbf{12}, 1--14 (2020).

\bibitem{13}
S.~Arnon and D.~Kedar, \enquote{Non-line-of-sight underwater optical wireless communication network,} {\protect\JournalTitle{J. Opt. Soc. Am. A}} \textbf{26}, 530--539 (2009).

\bibitem{14}
S.~Arnon, \enquote{{Underwater optical wireless communication network},} {\protect\JournalTitle{Optical Engineering}} \textbf{49}, 015001 (2010).

\bibitem{15}
F.~Jasman and R.~J. Green, \enquote{Monte carlo simulation for underwater optical wireless communications,} in \emph{2013 2nd International Workshop on Optical Wireless Communications (IWOW),}  (2013), pp. 113--117.

\bibitem{16}
S.~Yildiz, I.~Baglica, B.~Kebapci, \emph{et~al.}, \enquote{Reflector-aided underwater optical channel modeling.} {\protect\JournalTitle{Optics letters}} \textbf{47 20}, 5321--5324 (2022).

\bibitem{17}
S.~Tang, Y.~Dong, and X.~Zhang, \enquote{On path loss of nlos underwater wireless optical communication links,} in \emph{2013 MTS/IEEE OCEANS - Bergen,}  (2013), pp. 1--3.

\bibitem{18}
Y.~Dong, S.~Tang, and X.~Zhang, \enquote{Effect of random sea surface on downlink underwater wireless optical communications,} {\protect\JournalTitle{IEEE Communications Letters}} \textbf{17}, 2164--2167 (2013).

\bibitem{19}
M.~Sharifzadeh and M.~Ahmadirad, \enquote{Performance analysis of underwater wireless optical communication systems over a wide range of optical turbulence,} {\protect\JournalTitle{Optics Communications}} \textbf{427}, 609--616 (2018).

\bibitem{20}
M.~V. Jamali, J.~A. Salehi, and F.~Akhoundi, \enquote{Performance studies of underwater wireless optical communication systems with spatial diversity: Mimo scheme,} {\protect\JournalTitle{IEEE Transactions on Communications}} \textbf{65}, 1176--1192 (2017).

\bibitem{21}
C.~D. Mobley, \emph{Light and Water: Radiative Transfer in Natural Waters} (Academic Press, 1994).

\bibitem{22}
F.~Dong, L.~Xu, D.~Jiang, and T.~Zhang, \enquote{Monte-carlo-based impulse response modeling for underwater wireless optical communication,} {\protect\JournalTitle{Progress In Electromagnetics Research M}} \textbf{54}, 137--144 (2017).

\bibitem{23}
R.~W. Preisendorfer, \emph{Unpolarized irradiance reflectances and glitter patterns of random capillary waves on lakes and seas, by Monte Carlo simulation}, vol.~63 (US Department of Commerce, National Oceanic and Atmospheric Administration~…, 1985).

\bibitem{24}
S.~Xu, P.~Yue, and X.~Yi, \enquote{Non-line-of-sight multiple reflection underwater wireless optical communications channel model based on a capillary waves rough sea surface,} {\protect\JournalTitle{J. Opt. Soc. Am. A}} \textbf{40}, 1116--1127 (2023).

\bibitem{25}
C.~S. Cox and W.~H. Munk, \enquote{Statistics of the sea surface derived from sun glitter,} {\protect\JournalTitle{Journal of Marine Research}} \textbf{13}, 198--227 (1954).

\bibitem{26}
S.~Tang, X.~Zhang, and Y.~Dong, \enquote{Temporal statistics of irradiance in moving turbulent ocean,} in \emph{2013 MTS/IEEE OCEANS - Bergen,}  (2013), pp. 1--4.

\bibitem{27}
T.~J. Schulz, \enquote{Optimal beams for propagation through random media,} {\protect\JournalTitle{Optics letters}} \textbf{30}, 1093--1095 (2005).

\bibitem{28}
O.~Korotkova, N.~Farwell, and E.~Shchepakina, \enquote{Light scintillation in oceanic turbulence,} {\protect\JournalTitle{WAVES IN RANDOM AND COMPLEX MEDIA}} \textbf{22}, 260--266 (2012).

\bibitem{29}
D.~Xu, P.~Yue, X.~Yi, and J.~Liu, \enquote{Improvement of a monte-carlo-simulation-based turbulence-induced attenuation model for an underwater wireless optical communications channel,} {\protect\JournalTitle{JOSA A}} \textbf{39}, 1330--1342 (2022).

\bibitem{30}
R.~J. Drost, T.~J. Moore, and B.~M. Sadler, \enquote{Uv communications channel modeling incorporating multiple scattering interactions,} {\protect\JournalTitle{JOSA A}} \textbf{28}, 686--695 (2011).

\bibitem{31}
S.~Xu, P.~Yue, and D.~Xu, \enquote{Joint scattering and reflection impulse response modeling for non-line-of-sight underwater wireless optical links,} in \emph{2023 International Conference on Ubiquitous Communication (Ucom),}  (2023), pp. 243--247.

\end{thebibliography}

%%%%%%%%%% If preparing manually:
% \begin{thebibliography}{1}
% \newcommand{\enquote}[1]{``#1''}

% \bibitem{Zhang:14}
% Y.~Zhang, S.~Qiao, L.~Sun, Q.~W. Shi, W.~Huang, L.~Li, and Z.~Yang,
%   \enquote{Photoinduced active terahertz metamaterials with nanostructured
%   vanadium dioxide film deposited by sol-gel method,}
%   {\protect\JournalTitle{Optics Express}} \textbf{22}, 11070--11078 (2014).

% \bibitem{Optica}
% {Optica}, \enquote{{Optica Publishing Group},}
%   \url{http://www.opg.optica.org}.

% \bibitem{FORSTER2007}
% P.~Forster, V.~Ramaswamy, P.~Artaxo, T.~Bernsten, R.~Betts, D.~Fahey,
%   J.~Haywood, J.~Lean, D.~Lowe, G.~Myhre, J.~Nganga, R.~Prinn, G.~Raga,
%   M.~Schulz, and R.~V. Dorland, \enquote{Changes in atmospheric consituents and
%   in radiative forcing,} in \enquote{Climate Change 2007: The Physical Science
%   Basis. Contribution of Working Group 1 to the Fourth assesment report of
%   Intergovernmental Panel on Climate Change,}  S.~Solomon, D.~Qin, M.~Manning,
%   Z.~Chen, M.~Marquis, K.~B. Averyt, M.~Tignor, and H.~L. Miler, eds.
%   (Cambridge University Press, 2007).

% \end{thebibliography}

\end{document}